\documentclass[aps,prx,showpacs,amsmath,amssymb,amsfonts,lengthcheck,twocolumn,longbibliography,superscriptaddress]{revtex4-2}

\usepackage[T1]{fontenc}
\usepackage{graphicx}
\usepackage{physics}
\usepackage{amsthm}
\usepackage{amsmath}
\usepackage{verbatim}
\usepackage{dcolumn}% Align table columns on decimal point
\usepackage{url}
\usepackage{bm}% bold math
\usepackage{epsf}
\usepackage{algorithm}
\usepackage{algpseudocode}
\usepackage{booktabs}
\usepackage{longtable}
\usepackage{array}
\usepackage[none]{hyphenat}

\usepackage{color}
\usepackage[normalem]{ulem}
\usepackage{pgfplots}
\usepackage[colorlinks=true,citecolor=blue,linkcolor=blue,urlcolor=blue]{hyperref}
\usepackage{orcidlink}%
\definecolor{myvermillion}{HTML}{D55E00}
\definecolor{myyellow}{HTML}{F0E442}
\definecolor{mygreen}{HTML}{009E73}
\definecolor{mypurple}{HTML}{CC79A7}
\definecolor{myblue}{HTML}{0072B2}
\definecolor{myorange}{HTML}{E69F00}
\definecolor{mygray}{RGB}{153,153,153}

\pgfplotsset{compat=newest}

\newcommand{\bla}{bla\\bla\\bla\\bla\\bla}

\makeatletter
\newcommand{\currentfontsize}{\f@size pt}
\makeatother

\begin{document}

%\title{Seasonal-encoded Quantum Machine Learning}
\title{Quantum machine learning geospatial time-series by seasonal encoding}

\author{Devjyoti Tripathy\orcidlink{0009-0005-7067-1090}}
\email{dtripathy@umbc.edu}
    \affiliation{Department of Physics, University of Maryland, Baltimore County, Baltimore, MD 21250, USA}
    \affiliation{Quantum Science Institute, University of Maryland, Baltimore County, Baltimore, MD 21250, USA}

\author{Reece Robertson\orcidlink{0000-0003-1064-0012}}
    \affiliation{Department of Computer Science and Electrical Engineering, University of Maryland, Baltimore County, Baltimore, MD 21250, USA}
    \affiliation{Department of Physics, University of Maryland, Baltimore County, Baltimore, MD 21250, USA}
    \affiliation{Quantum Science Institute, University of Maryland, Baltimore County, Baltimore, MD 21250, USA}

\author{Josey Stevens\orcidlink{0000-0003-0627-1592}}
    \affiliation{Department of Physics, University of Maryland, Baltimore County, Baltimore, MD 21250, USA}
    \affiliation{Quantum Science Institute, University of Maryland, Baltimore County, Baltimore, MD 21250, USA}

\author{Catherine Nakalembe\orcidlink{0000-0002-2213-593X}}
\affiliation{Department of Geographical Sciences, University of Maryland, College Park, MD 20742, USA}

\author{Sebastian Deffner\orcidlink{0000-0003-0504-6932}}
    \affiliation{Department of Physics, University of Maryland, Baltimore County, Baltimore, MD 21250, USA}
    \affiliation{Quantum Science Institute, University of Maryland, Baltimore County, Baltimore, MD 21250, USA}
    \affiliation{National Quantum Laboratory, College Park, MD 20740, USA}
\date{\today}

%%%%%%%%%%%%%%%%%%
% ABSTRACT
%%%%%%%%%%%%%%%%%%
\begin{abstract}
Quantum data reuploading models can increase the expressive capacity of variational quantum models, but do not necessarily improve predictive performance. In the present work, we investigate this behavior for seasonal time series forecasting and show that data encoding determines how well temporal structures are represented as circuit depth increases. We develop a fidelity-based quantum geometric description in which repeated encodings contribute to an effective input generator, so that reuploading depth changes the quantum geometry through which approximately recurrent inputs are distinguished. We construct fixed and adaptive seasonal encodings that represent observation magnitude and seasonal phase in distinct Bloch-sphere coordinates. We evaluate these encodings for monthly precipitation derived from ERA5-Land, a global reanalysis dataset, from a representative semi-arid, drought-prone tropical region in East Africa, using a 14-month lag window and reuploading depths $L$= 1-6. We find that in ideal simulations, standard value-based encodings degrade markedly with depth, whereas adaptive seasonal encodings preserve temporal correlation, spectral structure, and forecasting accuracy. Noisy simulations show that this encoding hierarchy persists at shallow and intermediate depths, while larger depths are limited by circuit noise and optimization difficulty. The seasonal quantum models remain competitive with established classical forecasting methods (ARIMA, random forests, and LSTMs). These results identify structure-aware quantum data encoding as a general strategy for building deeper quantum reuploading models. 
\begin{comment}
    In ideal simulations, standard value-based encodings degrade markedly with depth, with test RMSE rising from [X] at L = 1 to \hl{[Y]} at L = 6, whereas adaptive seasonal encodings preserve temporal correlation, spectral structure, and forecasting accuracy, reducing test RMSE by [Z]\% relative to the best value-based encoding at L = [N]

    The seasonal quantum models remain competitive with established classical forecasting methods (ARIMA, random forests, and LSTMs), achieving \hl{[MAE/RMSE/R²]} of \hl{[values]} on the 105-month test set.
\end{comment}
\end{abstract}

\maketitle

%%%%%%%%%%%%%%%%%%
% INTRODUCTION
%%%%%%%%%%%%%%%%%%

\section{Introduction}

Remote sensing is now central to observing environmental changes across spatial and temporal scales. Optical, radar, thermal, and LiDAR sensors provide repeated measurements of, for instance, land cover and land-use change, forest structure, surface temperature, water availability, and precipitation \cite{Hansen2013globalforestchange, Wan2004, dubayah_global_2020}. These measurements are important not only for characterizing the present state of the Earth's environment, but also for understanding how it evolves.
% Precipitation time-series data are especially critical as they inform agricultural productivity, water resource availability, drought and flood conditions, and food security monitoring \cite{adebayo2026sub}.
Precipitation time-series data are especially critical for understanding agricultural productivity, water resource availability, drought and flood conditions, and food security monitoring \cite{adebayo2026sub, Shukla2021Harbinger, munoz-sabater_era5-land_2021}. However, the diversity of data modalities is growing and the resulting data volumes are increasing. Thus, developing novel methods for the extraction and forecasting of meaningful spatio-temporal patterns is an increasingly important problem in geospatial analysis.

Time series forecasting differs fundamentally from analyzing unordered observations, since predictive information is encoded not only in the values themselves, but also in their temporal organization. Environmental signals often exhibit recurring seasonal patterns, trends, and correlations across time scales, yet they are generally not perfectly periodic. Classical methods, from autoregressive models to random forests to recurrent neural networks, exploit these dependencies in different ways. Across these methods, performance depends on model capacity as well as on whether the representation preserves the relevant temporal relationships. As environmental forecasting problems become increasingly multivariate and spatiotemporal, the relevant predictive structure can involve correlations across time, and also across variables, spatial locations, and observational modalities. Therefore, a central challenge and opportunity is to determine whether new computational paradigms can offer representations that better preserve these predictive structures. %Here we focus on a univariate time series, monthly precipitation at a single location, which isolates the temporal problem before extending the approach to spatiotemporal data.

% by the European Center for Medium-Range Weather Forecasts (ECMWF)

Quantum machine learning (QML) provides a novel framework for constructing such representations by embedding classical data into quantum states and processing them with trainable quantum circuits \cite{prl-schuld-2019-qml-feature,
iop-Benedetti-2019-pqc-ml,
schuld_is_2022, nature-Schetakis-2022-review-qml, arxiv-kim-2025-qllm, nature-Biamonte-2017-qml-overview, nature-Huang-2021-power-data, iop-Guan-2021-qml-hep, PRXQ-larocca-2022-group-qml, nature-Caro-2022-qml-few-data, prxq-cincio-2026-qcnn-classical}. The ability of quantum feature maps and entangling operations to construct correlated representations makes QML particularly interesting for learning problems in which multiple interacting sources of structure must be represented simultaneously. Whether this representational flexibility translates into useful predictive performance remains an open question and should first be established in controlled settings where the relevant structure can be identified and interpreted. In variational quantum models, the choice of data encoding determines how relationships in the classical input space are translated into the geometry of Hilbert space and can therefore strongly influence what the model is able to learn \cite{nature-Havl-2019-qml-feature-space,
nature-Cerezo-2022-qml-challenge,
nature-schetakis-2025-qml-traffic, pra-hubregsten-2022-train-kernels, pra-elham-2026-qfeature-map}. Data reuploading architectures extend this idea by repeatedly inserting the classical input between trainable variational transformations, allowing increasingly rich nonlinear functions to be represented as the number of reuploading layers grows \cite{pra-schuld-2021-qdr-express,
pra-perez-2021-onequbit-approx,
quantum-perez-2020-qdr-univ-classif}. Increasing reuploading depth, however, does not necessarily lead to improved predictive performance and can be accompanied by challenges optimizing the model as the behavior of the objective function varies significantly  \cite{nature-McClean-2018-barren-plateau,
nature-Cerezo-2021-barren-plateau,
aqt-sim-2019-pqc-hybrid-express,
nature-Du-2022-circ-search}. Since the input is repeatedly re-encoded after successive quantum transformations, increasing depth also changes how the circuit represents relationships between different inputs. 

Understanding this depth-dependent representation is therefore particularly important for structured data such as time series, where predictive information depends on preserving meaningful relationships between temporally related observations. 
Before extending such models to multivariate and spatiotemporal settings, though, it is essential to establish how they behave in a controlled forecasting problem where the relevant temporal organization can be isolated. A univariate seasonal time series provides such a setting: the dominant recurrence seasonal structure can be identified directly, the effect of quantum data representation can be examined independently of cross-variable interactions, and forecasting performance can be compared against classical approaches. 

%Establishing this baseline is important before asking whether quantum representations can provide additional value in more complex multivariate settings.

Most commonly used quantum angle encodings map numerical input features directly through their magnitude into single-qubit rotations \cite{arxiv-sychen-2020-qlstm,
frontiers-Khan-2024-qlstm-solar,
nature-schetakis-2025-qml-traffic, 
epj-Moon-2025-rqnn, iop-Cao-2024-qutrit-qml, NEURIPS-yu-2024-error-bounds} while carrying no explicit information about where that observation occurs within a recurring temporal cycle. For a seasonal signal, these two pieces of information are qualitatively distinct. Similar precipitation values observed at different times of the year can have different temporal meanings, while observations separated by one annual cycle can occupy corresponding seasonal positions even when their magnitudes differ. A value-based encoding therefore requires the variational circuit to infer this organization indirectly from the repeated sequence of observations.
This distinction becomes especially relevant in data reuploading models where repeated encodings inserted between variational layers are known to modify the function classes accessible to a variational quantum model \cite{quantum-perez-2020-qdr-univ-classif,pra-schuld-2021-qdr-express}. Recent work in quantum machine learning has emphasized the value of incorporating known structure directly into quantum representations rather than relying only on unconstrained model capacity \cite{PRXQuantum-west-equiv-qml,
prr-li-2024-symm-qnn,
qmi-Heredge-2024-perm-inv-qnn}. This motivates a broader design question for quantum time series learning: when a physically or statistically meaningful temporal structure is already known, should that structure be incorporated directly into the quantum data encoding, or left for the variational circuit to infer?

In this work, we address this question by directly linking temporal recurrence to the geometry induced by a quantum data reuploading circuit. We consider input windows separated by a characteristic recurrence period and use their quantum-state fidelity to quantify how closely recurring observations are represented as reuploading depth increases. A local geometric analysis relates the distinguishability of these inputs to a Fubini-Study metric \cite{prl-braunstein-1996-qgeo,
quantum-stokes-2020-qng,
quantum-Spehner-2025-buresgeodesics, mathphys-provost-1980-reimannian} generated by contributions from each reuploading layer, showing explicitly that circuit depth changes the geometry through which pre-existing temporal differences in the data are resolved. As a result, we construct season-aware quantum encodings that represent observation magnitude and seasonal phase via separate coordinates on the Bloch sphere. Precipitation magnitude determines the polar coordinate, while position within the annual cycle or calendar month determines the azimuthal coordinate. We consider a fixed seasonal representation and its adaptive variants in which this reference geometry can be deformed during training, allowing the model to learn from the data without discarding the known temporal organization of the signal.

We demonstrate these ideas using monthly precipitation forecasting for Napak, Uganda, with a hybrid quantum data reuploading model across reuploading depths \(L=1,\dots,6\). To this end, we compare conventional value-based angle encodings with the fixed and adaptive seasonal encodings in ideal simulations, and subsequently examine the seasonal models under noisy simulations. We further benchmark the resulting forecasts against classical statistical and machine learning approaches: ARIMA (autoregressive integrated moving average) \cite{book-box-1970-arima}, random forest \cite{mlj-breiman-2001-rf}, and LSTM (long short-term memory) \cite{neuralcomp-Hochreiter-1997-lstm}. 

As a main result we find that in ideal simulations, the value-based encodings progressively lose forecasting performance and temporal structure as reuploading depth increases, whereas the adaptive seasonal encodings remain substantially more stable across depth. This hierarchy is also retained in noisy simulations at shallow and intermediate depths, but circuit noise and optimization difficulty begin to dominate at larger \(L\). Taken together, these results show that increasing circuit depth alone is insufficient to improve quantum time-series learning. Rather, the structure by which the data is encoded can play a central role in determining whether predictive temporal information remains accessible as the model deepens.

\begin{figure*}[t]
    \centering
    \begin{minipage}{\textwidth}
        \centering
        \includegraphics[width=\linewidth]{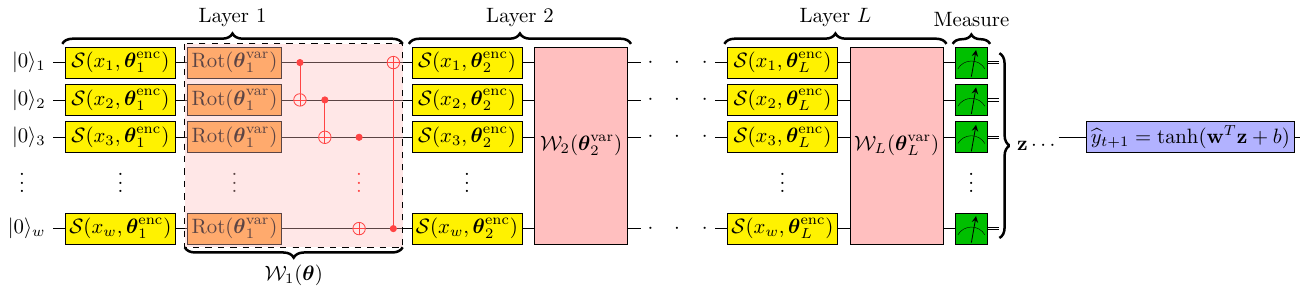}
        % \refstepcounter{figure}\label{fig:first}
        % \textbf{(\alph{figure})} Caption for the first image.
    \end{minipage}
    \vspace{1em}
    \begin{minipage}{\textwidth}
        \centering
        \includegraphics[width=\linewidth]{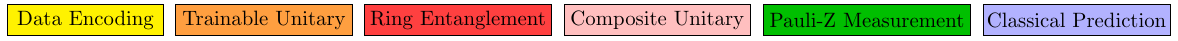}
        % \refstepcounter{figure}\label{fig:second}
        % \textbf{(\alph{figure})} Caption for the second image.
    \end{minipage}
    \caption{
        The hybrid quantum-classical machine learning model for time-series forecasting. Within each reupload layer, classically preprocessed data enters the quantum circuit through a data encoding block, followed by a variational block.  This is repeated $L$ times followed by Pauli-$Z$ measurement on all qubits, which is then passed onto a classical output head to obtain the prediction after classical post-processing.
        This circuit includes trainable parameters in the encoding step ($\boldsymbol\theta^{\text{enc}}_{\ell}$), which parameter is omitted in some of the encodings.
        Figure generated using $\langle\text{q}|\text{pic}\rangle$ \cite{Draper2016}.
    }
    \label{fig:hybrid-qc}
\end{figure*}

%%%%%%%%%%%%%%%%%%
% SETUP 
%%%%%%%%%%%%%%%%%%

\section{Hybrid quantum-classical machine learning}\label{sec:setup}

We begin by describing the hybrid quantum-classical forecasting model used throughout this work. Given a temporally ordered sequence of observations, the forecasting task is formulated as a one-step-ahead prediction from a lag window. We implement this map using a quantum data reuploading architecture in which the same input is repeatedly encoded between variational circuit blocks  \cite{quantum-perez-2020-qdr-univ-classif,quantum-Wach-2023-qdr-qubit,nature-schetakis-2025-qml-traffic, arxiv-wang-2025-qdr-predict,quantum-Barthe-2024-gradient-qdr}. This construction allows us to vary the reuploading depth while keeping the underlying forecasting task and readout architecture fixed.
\begin{comment}
Time series forecasting relies on extracting predictive structure from a sequence of temporally ordered observations. In a seasonal signal, this structure includes approximate recurrences; i.e., observations separated by a characteristic period can occupy similar positions without being exactly identical. In a quantum data reuploading model, the same input is encoded repeatedly between trainable quantum layers \cite{quantum-perez-2020-qdr-univ-classif,quantum-Wach-2023-qdr-qubit,nature-schetakis-2025-qml-traffic, arxiv-wang-2025-qdr-predict,quantum-Barthe-2024-gradient-qdr}. Increasing the number of reuploading layers changes not only the expressive capacity of the model, but also the quantum geometry through which temporally related inputs are represented. In this section, we formalize this connection. We first define the data reuploading forecasting model and then determine how the quantum distinguishability of recurrent input windows depends on encoding and reuploading depth.
\end{comment}

Consider a univariate time series
$\{y_0,y_1,\ldots,y_t,\ldots\}$. 
To predict the observation at time $t+1$, we construct a lag window containing the $w$ preceding observations
\begin{equation}
\mathbf{x}_t
=
\left(
y_{t-w+1},\ldots,y_{t-1},y_t
\right)^T  \in\mathbb{R}^{w},
\label{eq:lag_window}
\end{equation}
and learn the one-step forecasting map
\begin{equation}
f_{\boldsymbol{\theta}}:
\mathbf{x}_t
\longmapsto
\widehat{y}_{t+1},
\label{eq:forecasting_map}
\end{equation}
where $\widehat{y}_{t+1}$ is the prediction of the subsequent
observation and $\boldsymbol{\theta}$ denotes the complete set of trainable parameters of the hybrid quantum-classical model. We realize this map using a quantum data reuploading architecture. Each component of the lag window is associated with one qubit, resulting in a $w$-qubit register initialized in  $\ket{\mathbf{0}}\equiv \ket{0}^{\otimes w}$. We separate these parameters according to their role:
$
\boldsymbol{\theta}
=
\left(
\boldsymbol{\theta}^{\mathrm{enc}},
\boldsymbol{\theta}^{\mathrm{var}},
\boldsymbol{\theta}^{\mathrm{head}}
\right)
$. Here, $\boldsymbol{\theta}^{\mathrm{enc}}$ contains parameters of
a trainable encoding block, $\boldsymbol{\theta}^{\mathrm{var}}$ contains the
parameters of the variational block, and
$\boldsymbol{\theta}^{\mathrm{head}}$ contains the parameters of the
classical output head.

The quantum circuit consists of $L$ successive data reuploading layers, cf. Fig. \ref{fig:hybrid-qc} for a sketch. At layer
$\ell$, the input window is first mapped onto the quantum register via a data encoding block
$
S_{\ell}
\left(
\mathbf{x}_t,
\boldsymbol{\theta}^{\mathrm{enc}}_{\ell}
\right)
$ followed by a trainable variational circuit block
$
V_{\ell}
\left(
\boldsymbol{\theta}^{\mathrm{var}}_{\ell}
\right).
$ 

For our present purposes, we implement \(V_{\ell}\) using a hardware-efficient variational ansatz consisting of a trainable three-parameter rotation on every qubit followed by a nearest-neighbor ring of CNOT gates, where the three rotation angles on each qubit are independently trainable. The raw data values are suitably transformed via classical pre-processing such that they can be encoded into $S_{\ell}$. One complete reuploading block is, therefore,
\begin{equation}
A_{\ell}
\left(
\mathbf{x}_t,
\boldsymbol{\theta}^{\mathrm{enc}}_{\ell},
\boldsymbol{\theta}^{\mathrm{var}}_{\ell}
\right)
=
V_{\ell}
\left(
\boldsymbol{\theta}^{\mathrm{var}}_{\ell}
\right)
S_{\ell}
\left(
\mathbf{x}_t,
\boldsymbol{\theta}^{\mathrm{enc}}_{\ell}
\right),
\label{eq:reupload_block}
\end{equation}
and the depth-$L$ circuit becomes
\begin{align}
U_L
\left(
\mathbf{x}_t,
\boldsymbol{\theta}^{\mathrm{enc}},
\boldsymbol{\theta}^{\mathrm{var}}
\right)
&=
A_LA_{L-1}\cdots A_1 .
\label{eq:reupload_circuit}
\end{align}
Throughout this work, the reuploading depth $L$ denotes the number of
complete encoding-variational blocks. For fixed data encodings,
$
S_{\ell}$ contains no trainable encoding parameters, whereas for adaptive encodings, the parameters $\boldsymbol{\theta}^{\mathrm{enc}}$ are optimized jointly with the variational circuit.

The quantum state representing the lag window after $L$ reuploading layers then reads
\begin{equation}
\ket{\psi_{L}(\mathbf{x}_t)}
=
U_L
\left(
\mathbf{x}_t\right)
\ket{\mathbf{0}},
\label{eq:qnn_state}
\end{equation} where the dependence on trainable parameters has been suppressed for ease of notation. The quantum state is converted into a finite-dimensional classical vector by performing a Pauli-$Z$ measurement on all qubits $\{Z_q\}_{q=1}^{w}$, which is the output of the quantum model,
\begin{equation}
\mathbf{z}_{t}^{(L)}
=
\Big\{\bra{\psi_{L}(\mathbf{x}_t)}
Z_q
\ket{\psi_{L}(\mathbf{x}_t)}\Big\}_{q=1}^{w}.
\label{eq:quantum_features}
\end{equation}

\begin{figure*}
    \centering
    \includegraphics[width=\textwidth]{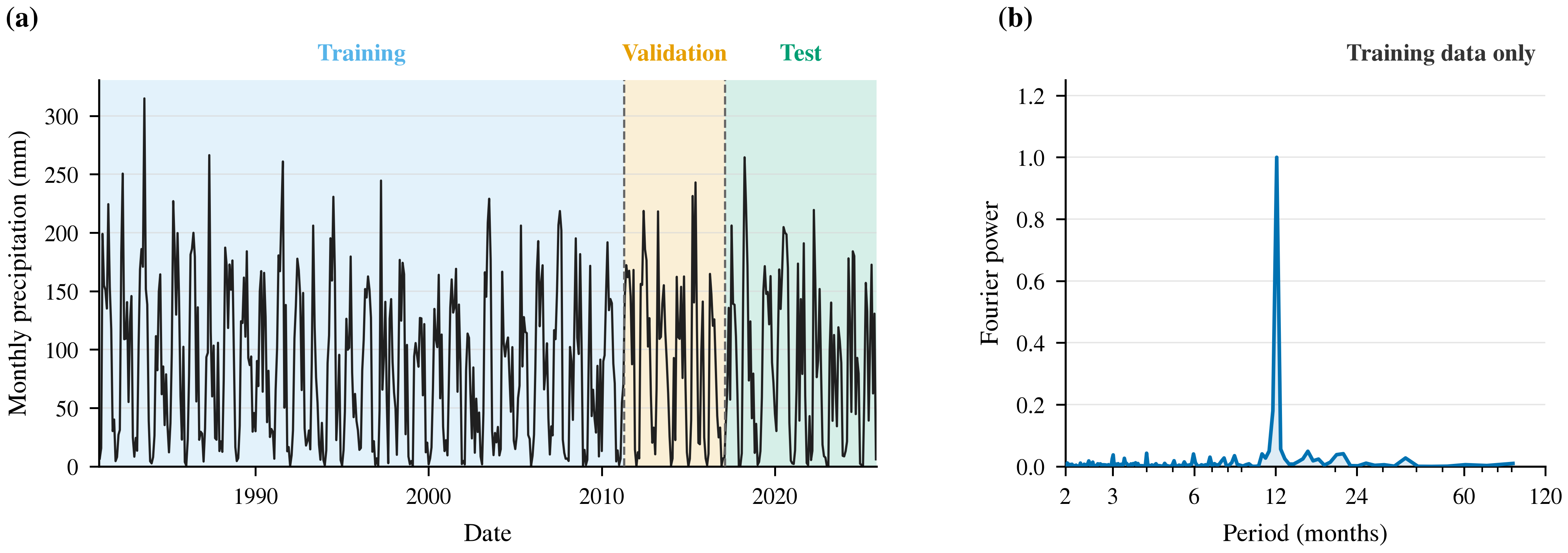}
    \caption{Dataset and seasonal structure
    \textbf{(a)} Monthly precipitation time series for Napak, partitioned chronologically into training, validation, and test intervals.
    \textbf{(b)} Normalized Fourier power of the training data only, computed after mean subtraction. The Fourier spectrum has a peak at approximately 12 months, indicating annual recurrence of the seasonal dataset.
    }
    \label{fig:dataset_spectrum}
\end{figure*}

These measurements are passed to a classical output head,
\begin{equation}
\widehat{y}_{t+1}
=
h_{\boldsymbol{\theta}^{\mathrm{head}}}
\left(\mathbf{z}_t^{(L)}\right)
=
\tanh
\left(
\mathbf{w}^{T}\mathbf{z}_t^{(L)}+b
\right),
\label{eq:classical_head}
\end{equation}
where $\widehat{y}_{t+1}$ denotes the forecast in the transformed target space, which is then transformed back into raw data units using classical post-processing. The forecasting model is trained by minimizing the empirical prediction loss over the training windows.

With the forecasting architecture established, we now turn to the precipitation time series on which it is evaluated. The temporal structure present in the data will provide the basis for examining how recurrent information is represented by the quantum model.

%%%%%%%%%%%%%%%%%%%%%%
% Dataset
%%%%%%%%%%%%%%%%%%%%%%%

\section{Dataset -- precipation time-series}\label{sec:dataset}

% \textcolor{red}{@Catherine: Could you please review this section on the dataset. I am not sure if this section needs references.}

% We study a monthly precipitation time series for Napak, Uganda spanning January 1981 through November 2025.

% \textcolor{red}{@Catherine: What is the source of the data, or how was it collected?}

We study a monthly precipitation time series for Napak, Uganda ($2.2^\circ\text{N}, 34.2^\circ\text{E}$), spanning January 1981 through November 2025. Napak is located in the drought-vulnerable Karamoja sub-region, an agro-pastoral area characterized by high climate variability and frequent, acute rainfall deficits~\cite{nakalembe_characterizing_2018}. Demonstrating the efficacy of quantum machine learning algorithms in a challenging region with such acute environmental dynamics establishes a robust baseline for extending the proposed architecture to similar climate-vulnerable regions globally. The monthly precipitation record was extracted from the ECMWF ERA5-Land Monthly Aggregated dataset (\texttt{ECMWF/ERA5\_LAND/MONTHLY\_AGGR}) via Google Earth Engine~\cite{munoz-sabater_era5-land_2021}. We retrieved the accumulated monthly precipitation (\texttt{total\_precipitation\_sum}) and converted the values from meters to millimeters.

The resulting record contains 539 consecutive monthly precipitation observations. We use a lag window of length $w=14$ months to predict precipitation in the subsequent month. Thus, each supervised example consists of 14 consecutive monthly observations followed by a one-month-ahead target. Figure~\ref{fig:dataset_spectrum}(a) shows the complete time series chronologically divided into 350 training targets spanning March 1982--April 2011, 70 validation targets spanning May 2011--February 2017 and 105 test targets spanning March 2017--November 2025. This chronological split preserves the temporal ordering of the forecasting problem while preventing leakage from future intervals into model training. 

The signal exhibits substantial month-to-month and year-to-year variability while retaining a recurrent annual structure. To characterize this recurrence, we compute the Fourier spectrum from the training data after subtracting the mean. As shown in Fig.~\ref{fig:dataset_spectrum}(b), the data exhibits a dominant annual frequency corresponding to a period of approximately 12 months. The data, however, is not perfectly periodic; lag windows separated by one annual cycle are generally similar but not identical. This motivates the central question considered next: how does a quantum data reuploading circuit distinguish such approximately recurrent input windows, and how does that distinction change as the reuploading depth increases?

\begin{comment}
We therefore identify the characteristic recurrence time as $T_0=12\ \mathrm{months}$. This timescale is then used to both compare quantum representations of seasonally related lag windows and to construct the seasonal quantum data encodings introduced next.
\end{comment}

%%%%%%%%%%%%%%%%%%%%%%%%%%%%%%
% FIDELITY THEORY
%%%%%%%%%%%%%%%%%%%%%%%%%%%%%%

\section{Theory -- ideal quantum circuit}\label{sec:theory}
Having shown that the time series has a characteristic recurrence period $T_0 = 12$ months and is approximately periodic, we ask how the quantum model represents two lag windows separated by one such period, $\mathbf{x}_t$ and $\mathbf{x}_{t+T_0}$. This comparison isolates the effect of repeated data encoding on temporally related inputs and provides the basis for analyzing how depth modifies the representation of seasonal structure. 

The similarity between two such inputs is measured by the fidelity,
\begin{equation}\label{eq:def_fidelity}
    F_L(t,t+T_0) = \abs{\langle\psi_{L}(\mathbf{x}_t)|\psi_{L}(\mathbf{x}_{t+T_0})\rangle}^2,
\end{equation}
with $0\leq F_L\leq1$. A value $F_L=1$ corresponds to identical quantum states up to a global phase, whereas smaller values indicate greater distinguishability in the quantum representation.
Suppose that the time series is perfectly periodic with period $T_0$, that is, $y_{n+T_0}=y_n\forall n $. Every component of the lag vector then repeats $\mathbf{x}_{t+T_0}
=
\mathbf{x}_t $. Since the trained circuit is a deterministic function of the input and trained parameters,
$
U_L(\mathbf{x}_{t+T_0})
=
U_L(\mathbf{x}_t)$ for arbitrary reuploading depth and arbitrary variational parameters. Consequently,

\begin{equation}
F_L(t,t+T_0)=1 \quad \forall L.
\label{eq:perfect_fidelity}
\end{equation}

Thus, repeated reuploading of the same data cannot distinguish two identical windows. Any depth-dependent change in the representation of seasonally related inputs must therefore originate from the fact that real seasonal signals are only approximately recurrent. 

We describe this imperfect recurrence by writing
\begin{equation}
\mathbf{x}_{t+T_0}
=
\mathbf{x}_t
+
\boldsymbol{\delta}_t ,
\label{eq:seasonal_delta}
\end{equation}
where $\boldsymbol{\delta}_t
=
\left(
\delta_{t,1},\ldots,\delta_{t,w}
\right)^T$ contains the cycle-to-cycle differences between corresponding components of the lag window. The displacement $\boldsymbol{\delta}_t$ is a property of the time series and exists before any input is processed by the quantum model. The role of the quantum circuit is therefore not to create this seasonal mismatch, but to determine how strongly this mismatch is represented in Hilbert space. 

To make this explicit, consider sufficiently small $\boldsymbol{\delta}_t$. The state associated with the displaced window can be expanded around $\mathbf{x}_t$ as
\begin{align}
\ket{\psi_L(\mathbf{x}_t+\boldsymbol{\delta}_t)}
=
\ket{\psi_L(\mathbf{x}_t)}
+
\sum_{j=1}^{w}
\delta_{t,j}
\partial_j
\ket{\psi_L(\mathbf{x}_t)}
+
\mathcal{O}(\|\boldsymbol{\delta}_t\|^2).
\label{eq:state_expansion}
\end{align}

The local distinguishability of two seasonally related windows is therefore controlled by the derivatives of the depth $L$ quantum state with respect to the input coordinates. Determining these derivatives requires accounting for the contribution of every reupload and the circuit transformations that precede it. Assume that the encoding is differentiable with respect to the chosen continuous input coordinates. For the $j^\text{th}$ component of the lag window at reuploading layer $\ell$, we define the encoding generator as
\begin{equation}
G_{j,\ell}
=
iS_\ell^\dagger(\mathbf{x}_t)
\partial_j S_\ell(\mathbf{x}_t),
\label{eq:bare_generator}
\end{equation}
so that
\begin{equation}
\partial_jS_\ell(\mathbf{x}_t)
=
-iS_\ell(\mathbf{x}_t)G_{j,\ell}.
\end{equation}

This generator describes the sensitivity of a single encoding layer to a change in input coordinate $x_j$. In a data-reuploading circuit, however, the same coordinate enters at every encoding layer, so the derivative of the complete circuit contains one contribution from each reupload. Define the partial circuits
\begin{equation}
B_\ell
=
A_\ell A_{\ell-1}\cdots A_1
\quad\text{with}\quad
B_0=I,
\label{eq:partial_circuit}
\end{equation}
so that $B_{\ell-1}$ contains all operations applied before the $\ell^\text{th}$ reupload. The resulting effective generator is
\begin{equation}
\mathcal{G}_j^{(L)}
\equiv
iU_L^\dagger\partial_jU_L
=
\sum_{\ell=1}^{L}
B_{\ell-1}^{\dagger}
G_{j,\ell}
B_{\ell-1}.
\label{eq:effective_generator}
\end{equation}

A straightforward calculation using Eq.~\eqref{eq:state_expansion} and Eq.~\eqref{eq:effective_generator} shows that the corresponding fidelity admits the local expansion
\begin{equation}
F_L(t,t+T_0)
=
1
-
\sum_{j,k=1}^{w}
g_{jk}^{(L)}
\delta_{t,j}\delta_{t,k}
+
\mathcal{O}(\|\boldsymbol{\delta}_t\|^3),
\label{eq:fidelity_metric}
\end{equation}
where $g^{(L)}$ is the Fubini-Study metric \cite{prl-braunstein-1996-qgeo, quantum-Spehner-2025-buresgeodesics} induced by the trained depth-$L$ circuit in input space. The metric can be written as the symmetrized covariance
\begin{equation}
g_{jk}^{(L)}
=
\frac{1}{2}
\left\langle
\left\{
\Delta\mathcal{G}_j^{(L)},
\Delta\mathcal{G}_k^{(L)}
\right\}
\right\rangle_0 .
\label{eq:metric_generator}
\end{equation}
where $\Delta\mathcal{G}_j^{(L)}
=
\mathcal{G}_j^{(L)}
-
\langle\mathcal{G}_j^{(L)}\rangle_0$ and $\langle O\rangle_0= \bra{\mathbf{0}}O\ket{\mathbf{0}}$. The structure is particularly transparent when projected directly onto the seasonal displacement $\boldsymbol{\delta}_t$. Define
\begin{equation}
G_{\boldsymbol{\delta},\ell}
=
\sum_{j=1}^{w}
\delta_{t,j}
G_{j,\ell}.
\end{equation}

Equation~(\ref{eq:fidelity_metric}) becomes
\begin{equation}
1-F_L(t,t+T_0)
\simeq
\operatorname{Var}_0
\left(
\sum_{\ell=1}^{L}
B_{\ell-1}^{\dagger}
G_{\boldsymbol{\delta},\ell}
B_{\ell-1}
\right),
\label{eq:fidelity_variance}
\end{equation}
where equality is understood to leading quadratic order in
$\boldsymbol{\delta}_t$. Equation~(\ref{eq:fidelity_metric}) separates two conceptually different contributions. The displacement $\boldsymbol{\delta}_t$ is a property of the data and exists before the quantum model is applied. The metric $g^{(L)}$, in contrast, describes how strongly the trained depth-$L$ circuit resolves that displacement. The details of the derivation are presented in the Supplemental Material.

The role of encoding is particularly transparent at a single reuploading layer. At $L=1$ the fidelitybecomes
\begin{align}
F_1(t,t+T_0)
&=
\left|
\bra{\mathbf{0}}
S_1^{\dagger}(\mathbf{x}_t)
V_1^{\dagger}V_1
S_1(\mathbf{x}_{t+T_0})
\ket{\mathbf{0}}
\right|^2
\nonumber\\
&=
\left|
\bra{\mathbf{0}}
S_1^{\dagger}(\mathbf{x}_t)
S_1(\mathbf{x}_{t+T_0})
\ket{\mathbf{0}}
\right|^2.
\label{eq:single_layer_fidelity}
\end{align}
The single reupload layer fidelity is therefore determined entirely by the encoding
map. Consequently, $F_1<1$ reflects the imperfect recurrence $\boldsymbol{\delta}_t$ already present in the data in the Hilbert space. At higher depths, intermediate variational blocks occur
between repeated encoding operations and rotate the layerwise
generators appearing in Eq.~\eqref{eq:effective_generator}, thereby changing the geometry of the temporal representation. 

For $L>1$, $F_L(t,t+T_0)<F_1(t,t+T_0)$ implies that the effective generator, which is transformed by all preceding encoding and variational layers through $B_\ell$, leads to degradation in the temporal representation in the quantum circuit. However, in general, multiple reupload depth may suppress, amplify, or redirect the pre-existing differences in the real time series. Importantly, Eqs.~(\ref{eq:fidelity_variance}) and \eqref{eq:fidelity_metric} do not imply that larger fidelity leads to better forecasting. A useful predictor must remain sensitive to genuine changes that contain predictive information. The result instead establishes that reuploading depth changes the geometry with which the circuit distinguishes temporally related inputs. Whether this change is beneficial depends on its alignment with the data's predictive structure. 

This observation provides the central motivation for the encoding strategies considered in this work. An encoding based only on the instantaneous value of an observation has no explicit information about where the observation occurs within a recurring temporal cycle. This suggests a general design principle for the quantum time series models: when a known temporal structure is predictive, it can be advantageous to encode the structure explicitly rather than requiring the variational circuit to infer it solely from repeated value-based reuploads. In the following sections, we apply this idea to construct quantum encodings that progressively incorporate and adapt to its seasonal organization.

\begin{figure*}
\centering
\includegraphics[width=\textwidth]
{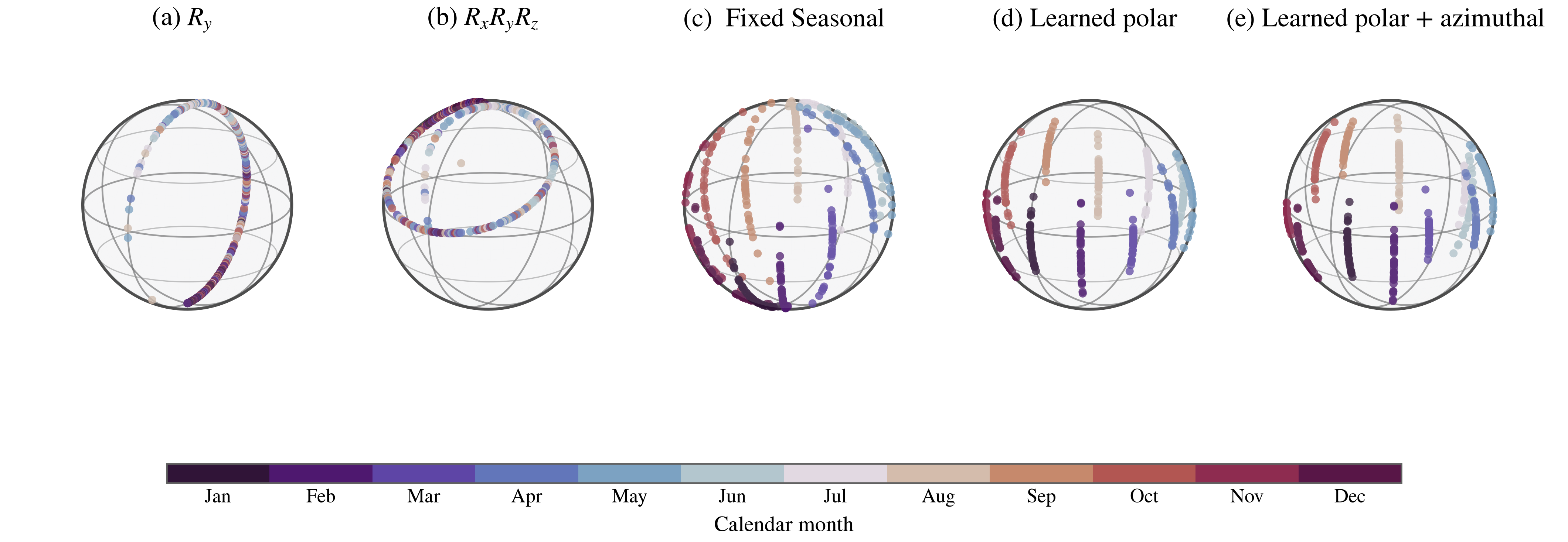}
\caption{Bloch-sphere geometry of the quantum data encodings
Each point represents one of the 364 monthly observations in the training set ( or equivalently, 350 training targets) and is colored according to calendar month.
\textbf{(a)} The $R_y$ encoding maps the scaled precipitation values onto a single meridian.
\textbf{(b)} The $R_xR_yR_z$ encoding applies the same data-dependent angle about all three rotation axes, producing a more intricate trajectory over the sphere.
\textbf{(c)} The fixed seasonal encoding maps the empirical training data's cumulative distribution to the polar coordinate and assigns the calendar month to the azimuthal coordinate. Overall, it maps the data onto 12 equally spaced seasonal meridians on the Bloch sphere.
\textbf{(d)} The learned polar encoding starts from the fixed seasonal encoding as the reference but later introduces month- and layer-dependent deformations of the polar mapping while retaining the fixed seasonal azimuths.
\textbf{(e)} The learned polar $+$ azimuthal encoding additionally permits bounded, month- and layer-dependent azimuthal adaptations in addition to the trainable polar coordinate. It also starts from the fixed seasonal encoding as the reference. For \textbf{(d)(e)}, the embeddings shown here are representative of a single layer from a single trained reuploading model; the learned polar mapping generally changes from layer to layer and from model to model.}

\label{fig:bloch_encodings}
\end{figure*}

%%%%%%%%%%%%%%%%%%%%%%
% Encodings
%%%%%%%%%%%%%%%%%%%%%%

\section{Quantum data encodings}
\label{sec:encodings}

The analysis in Sec.~\ref{sec:theory} showed that the geometry of approximately recurring inputs depends on the data encoding map, while Sec.~\ref{sec:dataset} identified a dominant annual frequency with $T_0=12$ months.  Therefore, we will compare a sequence of quantum encodings that progressively incorporate this temporal structure. 

The first two encodings depend only on precipitation magnitude and serve as value-based baselines. The remaining three encodings explicitly account for the position of each observation within the annual cycle, with the last two allowing this seasonal geometry to adapt during training. In all five cases, the lag windows, variational circuit, quantum readout, and classical prediction head are kept fixed, so that the differences between models arise from data representation itself.

For an input window
$\mathbf{x}_t
=
\left(
y_{t-w+1},\ldots,y_t
\right)^T$ qubit $q\in\{1,\ldots,w\}$ represents the observation at the index
\begin{equation}
r_{tq}=t-w+1+(q-1).
\label{eq:source_index}
\end{equation}
Since the time series has a characteristic period $T_0=12$, we associate each observation with the seasonal index 
\begin{equation}
    m_{tq} = r_{tq}\bmod T_0.
\end{equation}
This implies that
\begin{equation}
    m_{tq}\in{\{0,\dots,T_0-1\}}.
\end{equation}
Thus, $r_{tq}$ identifies the observation within the full time series, whereas $m_{tq}$ identifies its position within the recurring annual cycle.

At reuploading layer $\ell$, the complete encoding operation is constructed from single-qubit operations as
\begin{equation}
    \bigotimes_{q=0}^{w-1}S_{\ell,q}\left(y_{{r_{tq}}},m_{tq}\right).
\end{equation}
The dependence of $m_{tq}$ is absent for value-based encodings introduced below and becomes explicit only for seasonal encodings.

\subsection{Value based encodings}
We first consider two fixed encodings that depend only on the precipitation magnitude. Let $\widetilde{y_r}\in[-1,1]$ denote the linearly scaled precipitation value corresponding to the physical observation $y_r$. Define the associated rotation angle $\pi\widetilde{y}_{r}$. Since these encodings contain no trainable encoding parameters and are identical at every reuploading layer, we suppress the layer index $\ell$.

\subsubsection{$R_y$ encoding}
\label{subsubsec:ry_encoding}

The simplest encoding maps each scaled precipitation value to a
rotation about the $y$ axis. The complete window encoder is, therefore,
\begin{equation}
S^{(R_y)}
\left(
\mathbf{x}_t
\right)
=
\bigotimes_{q=0}^{w-1}
R_y
\left(
\pi\widetilde{y}_{r_{tq}}
\right).
\label{eq:ry_window_encoding}
\end{equation}
The encoded state is restricted to a single great circle around the Bloch sphere or, in other words, on a single Bloch sphere meridian as shown in Fig.~\ref{fig:bloch_encodings}(a). The magnitude of precipitation determines the position along this one-dimensional trajectory and contains no information about the seasonal position of the observation.

\subsubsection{$R_xR_yR_z$ encoding}
\label{subsubsec:rxryrz_encoding}

The second encoding applies the same data-dependent angle
about all three rotation axes. The corresponding
encoding operator is
\begin{equation}
S^{(R_xR_yR_z)}
\left(
\mathbf{x}_t
\right)
=
\bigotimes_{q=0}^{w-1}
R_z(\pi\widetilde{y}_{r_{tq}})
R_y(\pi\widetilde{y}_{r_{tq}})
R_x(\pi\widetilde{y}_{r_{tq}}).
\label{eq:rxryrz_window_encoding}
\end{equation}

This construction generates a more intricate trajectory than the
$R_y$ encoding, but the same scalar $\pi\widetilde{y}_{r_{tq}}$ controls all
three rotations as shown in Fig.~\ref{fig:bloch_encodings}(b). This keeps the representation one-dimensional with respect to the information supplied by each observation.

\subsection{Fixed seasonal CDF encoding}
\label{subsec:seasonal_cdf_encoding}

The $R_y$ and $R_xR_yR_z$ encodings introduced above distinguish observations only through their precipitation magnitude. To incorporate seasonal recurrence, we construct a two-dimensional representation in which the precipitation magnitude and the seasonal phase control two independent angular coordinates on the Bloch sphere.
We first map precipitation magnitude using the empirical cumulative distribution function derived from the training data. Let the empirical training CDF (cumulative distribution function) be denoted by $\mathrm{CDF}(y)$ as defined in Eq.~\eqref{eq:app-empirical-cdf} in the Supplemental Material. For the observation encoded on qubit $q$, define
\begin{equation}
u_{tq}
=
\mathrm{CDF}
\left(
y_{r_{tq}}
\right)\quad\text{and}
\quad
\zeta_{tq}
=
2u_{tq}-1,
\label{eq:cdf_to_z}
\end{equation}
where $\zeta_{tq}\in[-1,1]$. The corresponding polar angle is
\begin{equation}
\alpha_{tq}^{\mathrm{CDF}}
=
\arccos\left(\zeta_{tq}\right).
\label{eq:cdf_polar_angle}
\end{equation}
The polar coordinate is determined by the relative position of the precipitation value within the training distribution rather than by its absolute magnitude. The seasonal coordinate is determined by the position of the same observation within the annual cycle. 

Using the seasonal index $m_{tq}$ defined above, we assign the fixed azimuthal angle,
\begin{equation}
\phi_{tq}
=
\frac{2\pi m_{tq}}{T_0}.
\label{eq:fixed_seasonal_azimuth}
\end{equation}
For $T_0=12$, the annual cycle is therefore partitioned into 12 equally spaced azimuthal directions, or in other words, the data are encoded on 12 seasonal Bloch sphere meridians as shown in Fig.~\ref{fig:bloch_encodings}(c).
Correspondingly, the encoding operator becomes 
\begin{equation}
S^{(\mathrm{CDF})}
\left(
\mathbf{x}_t
\right)
= \bigotimes_{q=0}^{w-1}
R_z\left(\phi_{tq}^{}\right)
R_y\left(\alpha_{tq}^{\mathrm{CDF}}\right).
\label{eq:fixed_seasonal_operator}
\end{equation}

The precipitation controls the polar coordinate, while the calendar month controls the azimuthal coordinate. This construction embeds the known annual structure into the quantum representation before any variational transformation is applied. In contrast to $R_y$ and $R_xR_yR_z$ encodings, two observations with similar precipitation magnitude but different calendar months are no longer represented identically, and observations occurring at the same calendar month in different years lie on the same seasonal Bloch sphere meridians even when their magnitudes differ. 

\subsection{Adaptive seasonal encodings}

The fixed seasonal CDF  encoding imposes a physically interpretable geometry in which precipitation magnitude determines the polar coordinate and calendar month determines the azimuthal angle. Next, we allow this geometry to adapt during training while retaining the annual recurrence structure as the reference configuration. In both adaptive encodings, the trainable encoding parameters depend on the seasonal index $m_{tq}$ and on the reuploading layer $\ell$, so that different phases of the annual cycle may be represented differently at different reupload depths. 

\subsubsection{Learned polar seasonal encoding}

First, we allow the polar coordinate to vary while keeping the seasonal azimuthal coordinate fixed. The CDF derived reference polar angle  $\alpha_{tq}^{(\mathrm{CDF})}$ is transformed through a month and layer dependent map $\alpha_{tq,\ell}^{(LP)}=\mathcal{A}_{m_{tq},\ell}
\left(
\alpha_{tq}^{(\mathrm{CDF})}
\right)$ where $\mathcal{A}_{m,\ell}$ denotes a trainable deformation of the polar coordinate. The seasonal azimuthal angle remains fixed as $\phi_{tq}$ so that the seasonal information is preserved. 

The encoding at reuploading layer $\ell$ is, therefore,
\begin{equation}
S_\ell^{\mathrm{(LP)}}(\mathbf{x}_t)=\bigotimes_{q=0}^{w-1}
R_z\left(
\phi_{tq}
\right)
R_y\left(
\alpha_{tq,\ell}^{(\mathrm{LP})}
\right).
\end{equation}
The trainable map is initialized to the identity, so it reduces to the fixed seasonal CDF encoding before training. Optimization subsequently allows the model to reshape the precipitation-dependent polar coordinate independently for each seasonal phase and reuploading layer, while preserving the fixed cyclic organization in azimuthal coordinate as shown in Fig.~\ref{fig:bloch_encodings}(d).

\subsubsection{Learned polar $+$ azimuthal seasonal encoding}
\label{subsec:learned_polar_azimuth_encoding}

The final encoding also allows the azimuthal coordinate to adapt by a trainable month- and layer-dependent map $\phi_{tq} + \Delta\phi_{m_{tq},\ell}$. The polar coordinate is adapted the same way as in the learned polar seasonal encoding. The corresponding encoding operator is
\begin{equation}
S_\ell^{\mathrm{(LPA)}}(\mathbf{x}_t)=\bigotimes_{q=0}^{w-1}
R_z\left(
\phi_{tq,\ell}^{(\mathrm{LPA})}
\right)
R_y\left(
\alpha_{tq,\ell}^{(\mathrm{LPA})}
\right).
\end{equation}

The azimuthal displacement $\Delta\phi_{m,\ell}$ is bounded so that the model can locally adjust the seasonal phase associated with each month without arbitrarily destroying the cyclic ordering of calendar months as shown in Fig.~\ref{fig:bloch_encodings}(e). The trainable map is initialized such that the encoding reduces exactly to the fixed seasonal CDF encoding.
The three encodings consequently form a nested family. The fixed seasonal CDF encoding imposes the reference geometry.

The value-based encodings in Fig.~\ref{fig:bloch_encodings}(a),(b) map the time-series data exclusively according to their precipitation magnitude and completely ignore their seasonal structure. The seasonal encodings in Fig.~\ref{fig:bloch_encodings}(c),(d),(e) take the seasonality into account and map the data onto 12 different meridians on the Bloch sphere. The adaptive variants deform this seasonal geometry through their learned encoding parameters. In both adaptive encodings, the encoding parameters are optimized jointly with other trainable parameters of the forecasting model. The details of the various encoding schemes are presented in the Supplemental Material.

\begin{figure*}
    \centering
    \includegraphics{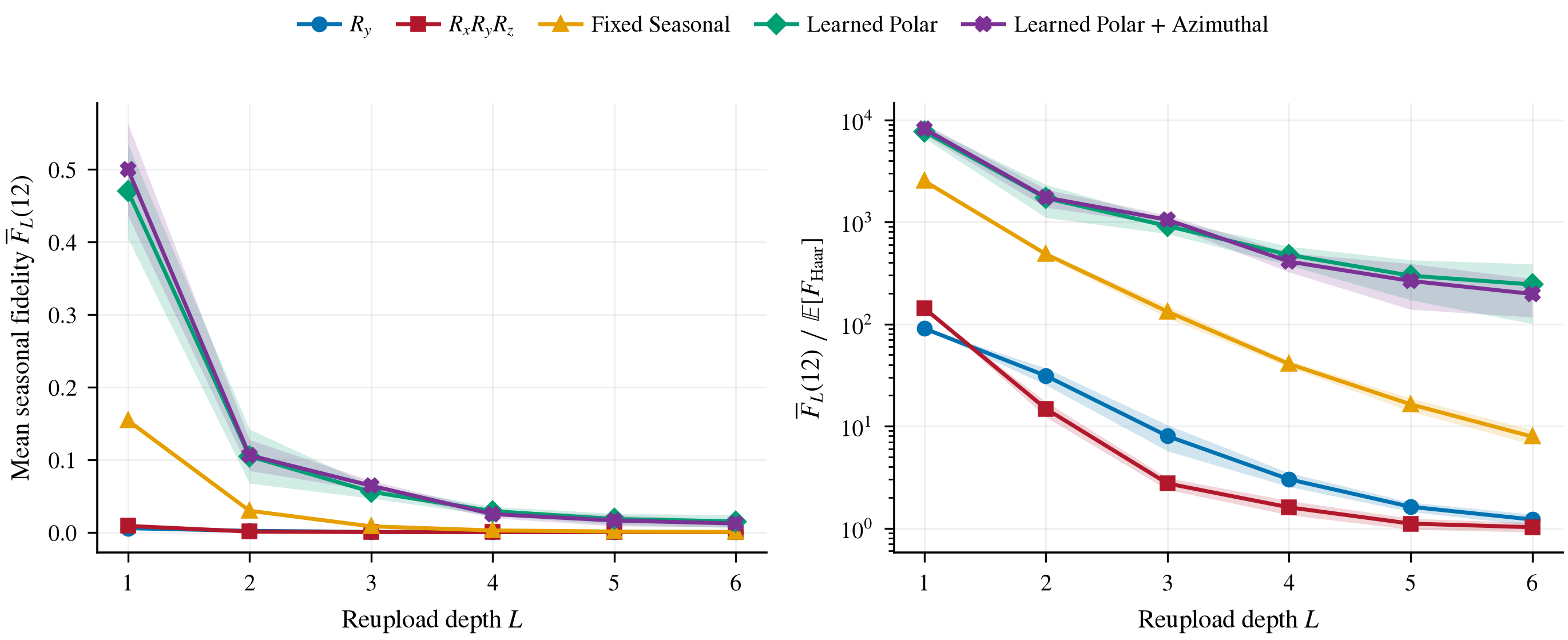}
    \caption{Same-season quantum fidelity as a function of reuploading depth $L$. (a) Mean seasonal fidelity $\overline{F}_L(12)$ measures the similarity between two quantum representations of two lag windows separated by 12 months, averaged over all possible pairs of lag windows in the test set. (b) The same fidelity normalized by expected fidelity between Haar-random states, highlighting similarity compared to a random state reference. Curves show the mean across five independently trained seeds, with shaded region indicating $\pm$ standard deviation. }
    \label{fig:fid-depth}
\end{figure*}

%%%%%%%%%%%%%%%%%%
% RESULTS
%%%%%%%%%%%%%%%%%%

\section{Results}
\label{sec:results}

We evaluate the five encodings within the same $ 14$-qubit data reuploading forecasting architecture, varying the number of reuploading layers over
$L=1,\ldots,6$.
For each encoding and depth, five independent models are trained, and
model selection is performed exclusively on the validation set.
The checkpoint with the lowest validation loss is retained for each run
and evaluated once on the test set.

Forecasting performance is assessed using four complementary quantities:
the test root mean squared error (RMSE) (Eq.~\eqref{eq:test-rmse}), which measures absolute prediction
error in millimeters, the Pearson correlation (Eq.~\eqref{eq:pearson-correlation}), which measures temporal alignment between the observed and predicted series, the Fourier power cosine similarity $C_{\mathrm{FFT}}$ (Eq.~\eqref{eq:fft-cosine}), which
compares their normalized frequency domain structure; and the Amplitude variance ratio (Eq.~\eqref{eq:variance-ratio}), which measures recovery of the observed forecast amplitude. The detailed definitions of these metrics are given in the Supplementary Material.

\begin{figure*}
\centering
\includegraphics{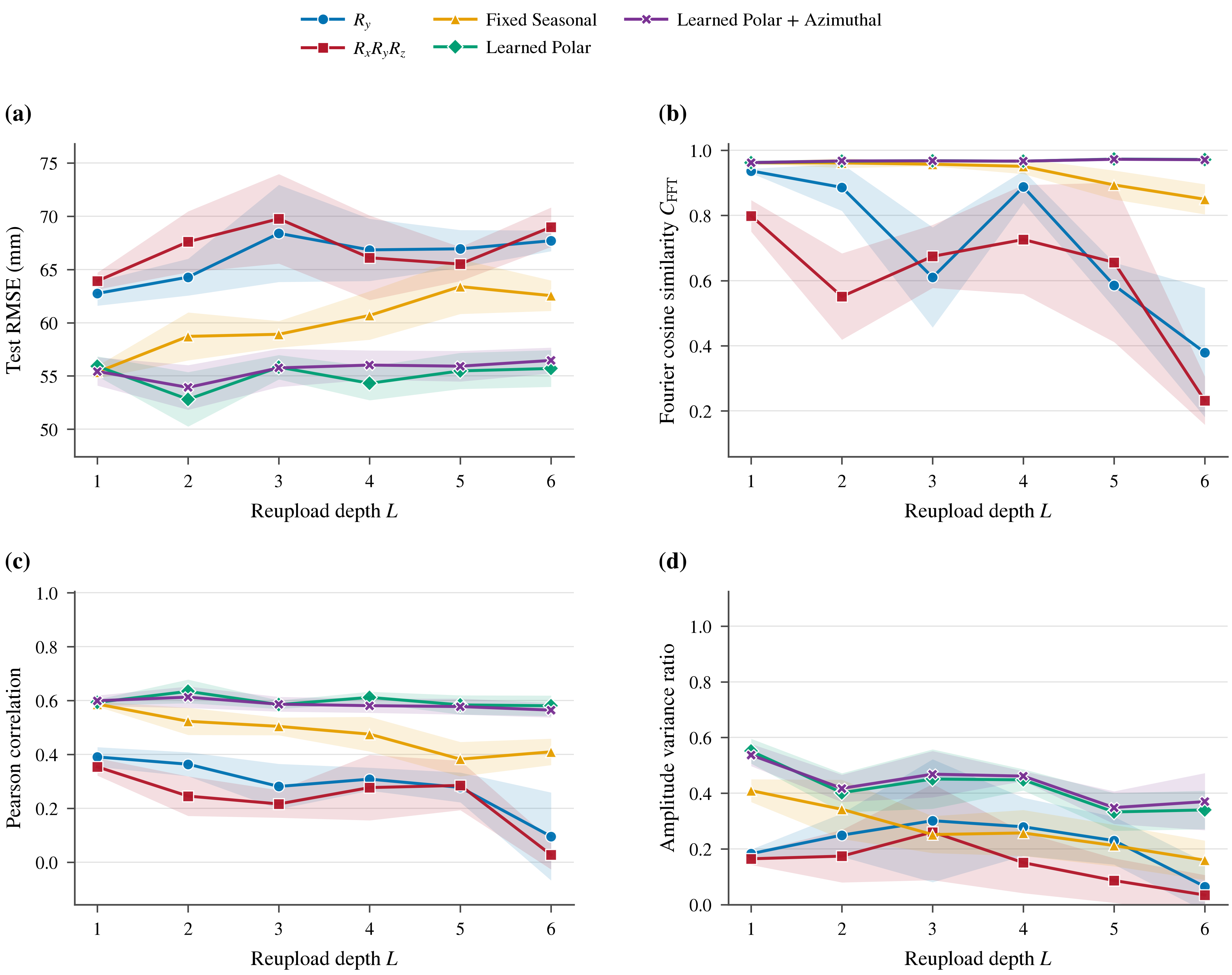}
\caption{Forecasting metrics shown as a function of reuploading depth $L$ across the five different encodings. (a)  Test root mean square error (RMSE) measures the magnitude of pointwise forecast error in precipitation units (mm). Lower values indicate better predictive accuracy. (b) Fourier cosine similarity measures the similarity between the normalized Fourier power spectra of the predicted and observed time series, thereby probing how well the forecast preserves dominant temporal frequencies. (c) Pearson correlation measures the linear correlation between the predicted and observed time series and captures whether temporal fluctuations are tracked correctly, independent of the absolute scale. (d) Amplitude variance ratio assesses whether the forecast reproduces the observed variability in amplitude. A value of 1 indicates matched variance. Curves denote the mean over five independently trained seeds, and shaded regions indicate $\pm 1$ standard deviation.}
\label{fig:ideal-metrics}
\end{figure*}

%%%%%%%%%%%%%%%%%%
% IDEAL SIMULATIONS
%%%%%%%%%%%%%%%%%%

\subsection{Ideal simulations}
\label{subsec:ideal_results}

The ideal simulations were performed using the \texttt{default.qubit} state-vector simulator and automatic differentiation framework by PennyLane \cite{bergholm-2022-pennylane-automaticdifferentiationhybrid}. The complete hybrid quantum-classical learning pipeline was implemented using PyTorch \cite{NEURIPS-paszke-2019-pytorch}.

\subsubsection{Representation of seasonal recurrence with depth}
\label{subsubsec:seasonal_fidelity}

We first examine how the different encodings represent inputs
separated by the dominant seasonal period.
This tests the quantum geometric picture developed in Sec.~\ref{sec:theory} before considering forecasting performance.

For a depth-$L$ circuit, we define the mean seasonal fidelity over all
valid test set window pairs separated by $T_0$ as
\begin{equation}
\overline{F}_L(T_0)
=
\frac{1}{N_{T_0,\mathrm{test}}}
\sum_t
\left|
\left\langle
\psi_{L}(\mathbf{x}_t)
\middle|
\psi_{L}(\mathbf{x}_{t+T_0})
\right\rangle
\right|^2,
\label{eq:mean_seasonal_fidelity}
\end{equation}
where $N_{T_0,\mathrm{test}}$ is the number of valid pairs in the test data. For our time series, $T_0=12~\mathrm{months}$. Thus, $\overline{F}_L(T_0)$ measures the average similarity between
quantum representations of lag windows separated by one annual cycle.

Figure~\ref{fig:fid-depth}(a) shows that the mean seasonal fidelity decreases
with reuploading depth for all five encodings. The rate and scale of this decrease, however, depend strongly on how the data are represented. The $R_y$ and $R_xR_yR_z$ encodings begin with relatively small seasonal overlaps and decrease rapidly with depth. By contrast, the three seasonal encodings produce substantially larger overlaps at shallow depth but also exhibit a pronounced decline with $L$.

The single-layer behavior is consistent with the geometric analysis of
Sec.~\ref{sec:theory}. For fixed encodings, the variational block cancels exactly from the fidelity at $L=1$, so that the overlap is determined entirely by the encoding map. Correspondingly, the fixed seasonal CDF representation gives a substantially larger single-layer seasonal overlap than either value-based encoding.
The adaptive encodings are initialized from the same seasonal geometry but subsequently optimize their encoding parameters during training, allowing the learned representation itself to change.

To interpret the magnitude of these fidelities in context, we compare them
with the characteristic fidelity of two independent Haar-random pure states \cite{pra-sommers-2005-haar-fidelity,aqt-Sim-2019-haar-qc} in a Hilbert space of dimension $d=2^{14}$. For such states, $\mathbb{E}\left[F_{\mathrm{Haar}}\right]=1/d$. We consider the normalized quantity $\overline{F}_L(12)/\mathbb{E}[F_{\mathrm{Haar}}]=d\,\overline{F}_L(12)$ shown in Fig.~\ref{fig:fid-depth}(b). At $L=6$, the $R_y$ and $R_xR_yR_z$ encodings have decreased to the $F_{\mathrm{Haar}}$ scale. The fixed seasonal CDF representation remains above $F_{\mathrm{Haar}}$, better than the value-based encodings. In contrast, the two adaptive seasonal encodings remain roughly two orders of magnitude above the Haar reference even at $L=6$. These results show that repeated data reuploading progressively separates the quantum representations of input windows related to different years, but the extent of this separation depends strongly on the encoding geometry.

Importantly, $\overline{F}_L(12)$ is a diagnostic of representation geometry rather than a forecasting metric. A reduction in fidelity does not by itself imply a loss of predictive information, since the seasonal differences between $\mathbf{x}_t$ and $\mathbf{x}_{t+T_0}$ may themselves be informative. We therefore next examine whether the depth-dependent changes in quantum geometry are accompanied by corresponding changes in forecasting performance.

\begin{figure*}
    \centering
    \includegraphics{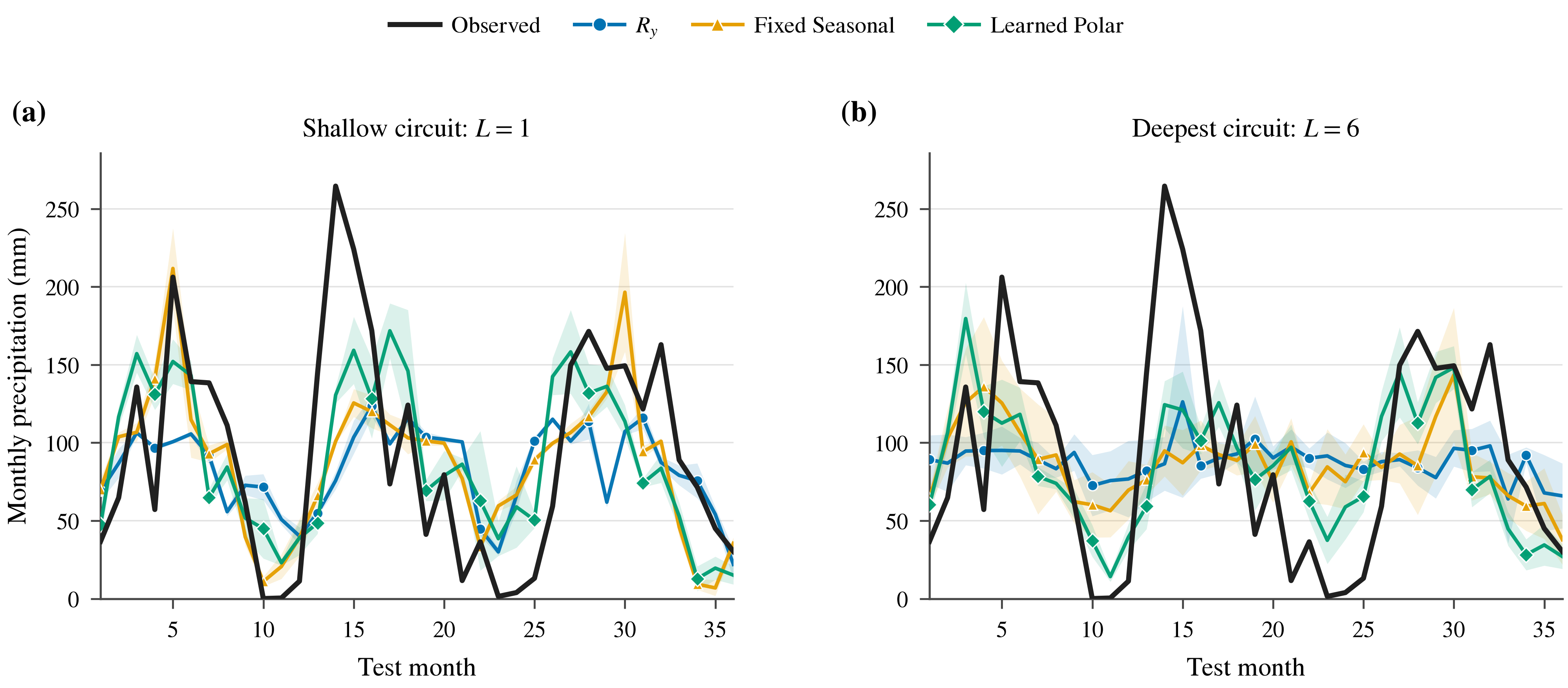}
    \caption{Observed monthly precipitation is compared with the mean forecast across five independently trained seeds for the $R_y$, fixed seasonal, and learned polar encodings over the first 36 months of the test data. (a) Shallow reuploading circuit with $L=1$. (b) Deepest reuploading circuit with $L=6$. At $L=6$, the learned polar encoding continues to retain good forecasts, the fixed seasonal encoding performs slightly better, and the $R_y$ encoding shows the poorest forecast quality.}
    \label{fig:36-month-forecast}
\end{figure*}

\subsubsection{Forecasting performance}
\label{subsubsec:ideal_forecasting}

Figure~\ref{fig:ideal-metrics} summarizes the test data predictions as a function of reuploading depth $L$. Figure~\ref{fig:ideal-metrics}(a) shows a clear separation between the seasonal and value-based encodings in terms of prediction error. The $R_y$ and $R_xR_yR_z$ encodings attain their lowest RMSE at shallow depth and generally deteriorate as additional reuploading layers are introduced. The fixed seasonal CDF encoding substantially improves upon these baselines at small depth, but its error also increases progressively with $L$. In contrast, both learned seasonal encodings remain comparatively stable across the full depth range. The lowest test error is obtained by the learned polar encoding at $L=2$, Test RMSE $ = 52.81\pm2.55~\mathrm{mm}$, whereas the best value-based encodings remain near $63~\mathrm{mm}$. 

A similar hierarchy appears in the frequency-domain results of Fig.~\ref{fig:ideal-metrics}(b). The learned seasonal encodings preserve an almost unchanged spectral structure across the complete depth range, with $C_{\mathrm{FFT}}\simeq 0.96\text{-}0.97$. The fixed seasonal CDF encoding also retains strong spectral agreement at shallow and intermediate depth, although its similarity decreases at the largest $L$. By comparison, both value-based encodings exhibit substantially stronger depth-dependent fluctuations and ultimately lose much of their agreement with the observed spectrum. 

The Pearson correlations in Fig.~\ref{fig:ideal-metrics}(c) show an even clearer difference in depth dependence. Both learned seasonal encodings maintain strong and comparatively stable temporal alignment throughout the entire range of depths. The learned polar encodings reach their largest correlation $0.634\pm0.044$ at $L=2$, and remain close to this level at larger depths. The learned polar $+$ azimuthal encoding exhibits a similar trend. The fixed seasonal encoding shows a more gradual loss of correlation with depth, whereas the value-based encodings deteriorate much more strongly and approach negligible temporal correlation by $L=6$.  Both learnable seasonal models show a slight increase in performance at $L=2$ that is not seen in any other model.  This highlights the distinction between predictive model performance and the seasonal fidelity of Eq.~(\ref{eq:mean_seasonal_fidelity}) and indicates that the optimization is able to effectively utilize some of the enhanced expresive capacity from data reuploading.

The amplitude variance ratios in Fig.~\ref{fig:ideal-metrics}(d) reveal a complementary limitation. All five models underestimate the amplitude of the observed precipitation fluctuations throughout the depth range. The learned seasonal models nevertheless retain substantially more variability than the value-based encodings and exhibit much weaker amplitude collapse as the circuit becomes deeper. The largest amplitude recovery occurs for the learned polar model at $L=1$.

Taken together, the four metrics show that the learned seasonal encodings are not only distinguished by a lower pointwise error, but also preserve temporal correlation, normalized spectral structure, and forecast amplitude much more consistently as reuploading depth increases. The fixed seasonal CDF representation captures a substantial fraction of this improvement at shallow depth, but its performance degrades more strongly with $L$, indicating that explicitly encoding the annual cycle is beneficial while allowing the corresponding geometry to adapt provides additional robustness.

The comparison with the fidelity results further shows that forecasting performance cannot be identified with preservation of seasonal recurrence alone. The seasonal fidelity of the learned encodings decreases substantially with depth, yet their Pearson correlation and Fourier similarity remain nearly unchanged. The learned seasonal representations therefore continue to retain forecast-relevant temporal structure even as annually separated input windows become increasingly distinguishable in Hilbert space. By contrast, for the value-based encodings, the reduction in seasonal fidelity is accompanied by a significant loss of temporal and spectral forecasting structure.

Finally, allowing the azimuthal coordinate to become trainable does not produce a consistent improvement over adapting the polar coordinate alone. The learned polar and learned polar $+$ azimuthal encodings exhibit very similar behavior across all four forecasting metrics, while the learned polar encoding attains the overall minimum RMSE and maximum Pearson correlation. For this time series, the additional azimuthal degrees of freedom therefore provide no predictive advantage beyond the learned polar representation and that the evenly-spaced seasonal encoding is near optimal.

\subsection{Noisy Simulations}
The ideal simulations show that the seasonal encodings remain comparatively robust as the number of reuploading layers increases. They further establish a clear hierarchy among the seasonal encodings, with the learned seasonal models generally outperforming the fixed seasonal CDF representation and remaining substantially more robust as reuploading depth increases. We next examine if this behavior persists in the presence of noise. For the noisy simulations, we evaluated the three seasonal encoding strategies: Fixed seasonal CDF, learned polar, and learned polar $+$ azimuthal across $L=1,\dots,6$.

\begin{figure}
    \centering
    \includegraphics[width=0.5\textwidth]{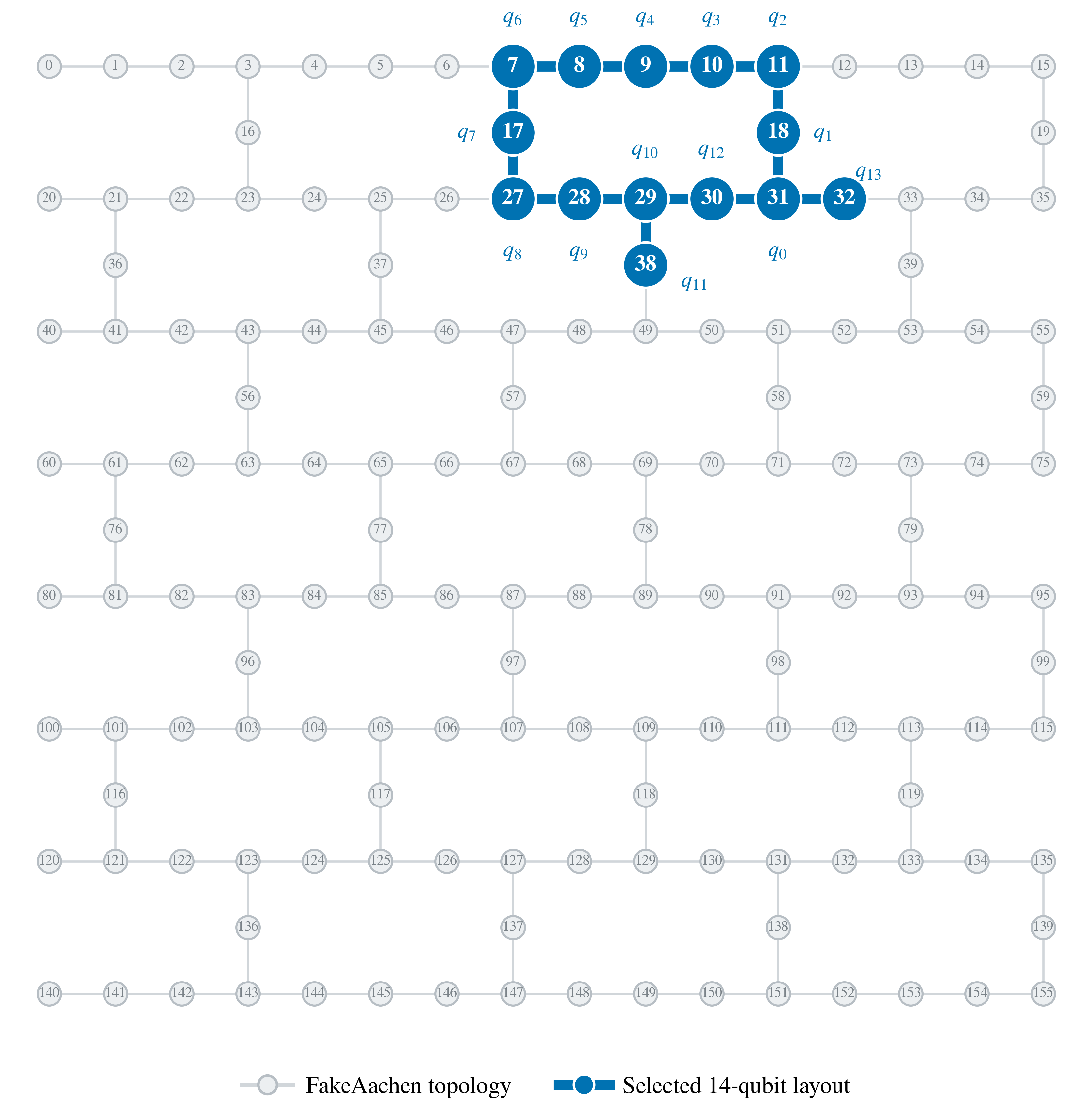}
    \caption{Topology of IBM FakeAachen used for noisy simulations. The 14-qubit subset selected for implementing the quantum reuploading circuit is highlighted, while the remaining qubits and couplings are shown in gray.}
    \label{fig:aachen-topology}
\end{figure}

Noisy circuit evaluations were performed with Qiskit Aer using the IBM FakeAachen noise model \cite{arxic-abhari-2024-qiskit}. Circuit executions used 512 measurement shots. The logical circuits were compiled onto a fixed 14-qubit physical layout as shown in Fig.~\ref{fig:aachen-topology}. The entire forecasting model was trained using a hybrid optimization procedure in which quantum circuit parameters were updated using simultaneous perturbation stochastic approximation (SPSA) \cite{ieee-spall-1992-spsa}, while classical parameters were optimized with Adam \cite{iclr-kingma-2015-adam}. To make the noisy simulations computationally tractable, the circuit evaluations were parallelized across multiple CPU cores on an HPC cluster. The details of the exact implementation are presented in the Supplemental Material.

\begin{figure*}
    \centering
    \includegraphics{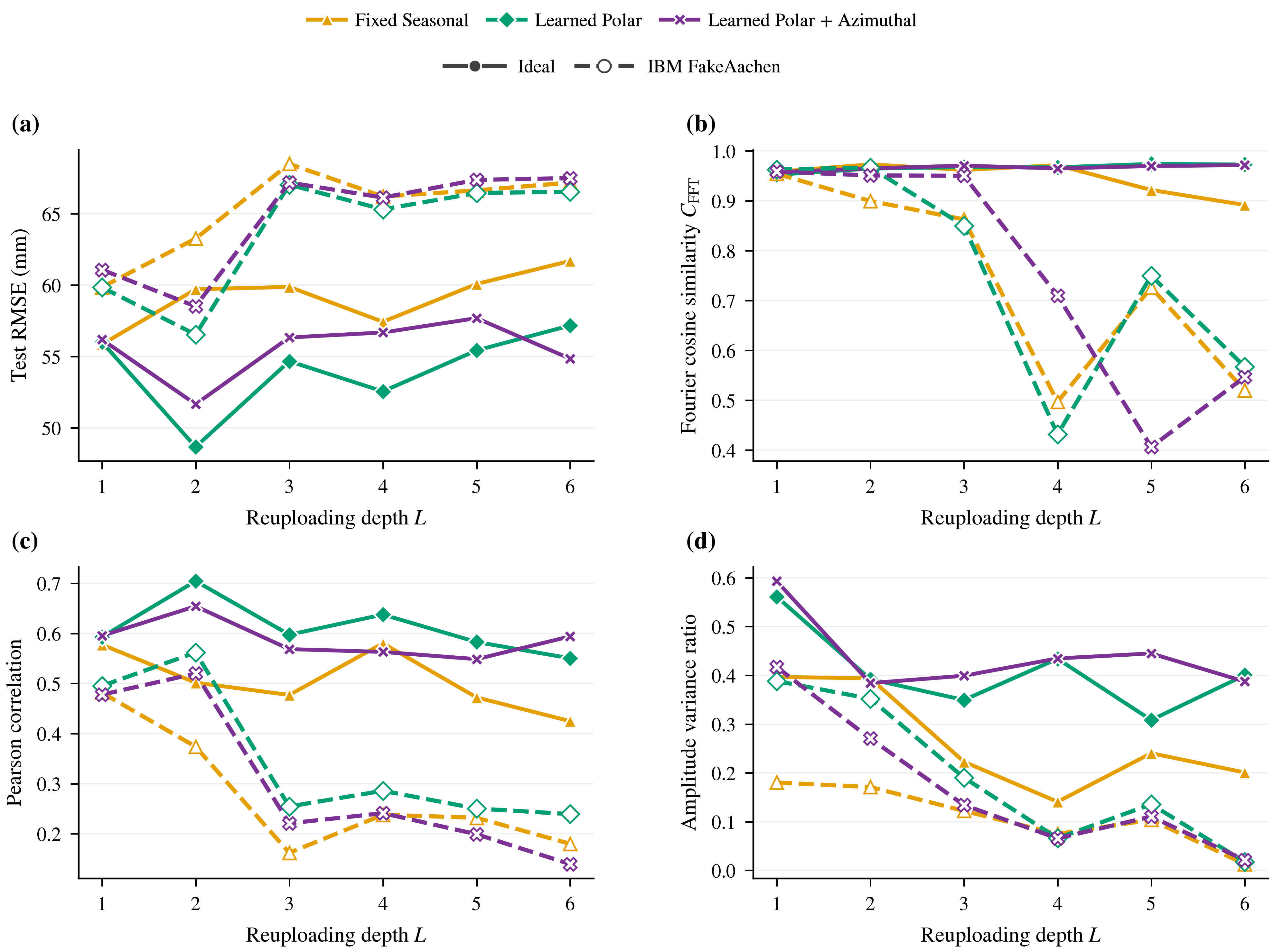}
    \caption{Comparison of ideal and noisy quantum simulations across reuploading depth \(L\) for the three seasonal encoding schemes. Both regimes are shown for a single matched initialization, rather than as an ensemble average across seeds. Solid lines with filled markers denote ideal simulations, while dashed lines with open markers denote simulations using the IBM FakeAachen noise model.}
    \label{fig:ideal-vs-noisy}
\end{figure*}

Figure~\ref{fig:ideal-vs-noisy} compares the noisy simulations with
their ideal counterparts over reuploading depths
$L=1,\ldots,6$. A central observation is that, at shallow and intermediate depths
$L=1,2,3$, the  hierarchy of encodings established in the ideal
simulations is preserved under noise. In particular, the adaptive seasonal encodings retain their advantage over the fixed seasonal CDF encoding across all the forecasting metrics. Thus, the benefit of learning the seasonal encoding geometry is not restricted to ideal state-vector simulations, but is also retained in the presence of hardware-derived noise. At these depths, hardware noise degrades the absolute forecasting performance relative to the ideal case, but it does not immediately erase the relative advantage produced by the encoding choice.

The situation changes as additional reuploading layers are introduced.
We again observe that all adaptive seasonal encoding models perform optimally at $L=2$, an observation that is not true of the fixed seasonal encodings.  
Beyond $L=3$, the forecasting metrics deteriorate substantially for all three encodings and the differences between them become progressively compressed. The test RMSE increases and eventually enters a similar high-error regime for the three seasonal models, while the Pearson correlation and forecast amplitude decrease more strongly with depth. By the deepest circuits, the predictions retain little temporal correlation with the observed precipitation series and exhibit severe amplitude suppression and spectral dissimilarity in the Fourier domain. These trends indicate that the advantages associated with the different
seasonal geometries become increasingly difficult to exploit as the depth and noise grow.

The degradation in performance at higher depths should also be interpreted in the context of the hybrid optimization procedure. During the noisy simulations, we find that convergence is sensitive to
the SPSA hyperparameters, particularly the parameter controlling the gradient updates \cite{ieee-spall-1998-spsa-a}. This sensitivity becomes increasingly relevant as the circuit depth grows and the stochastic optimization landscape becomes more difficult to navigate. Consequently, the deterioration observed at large $L$ need not arise solely from the accumulation of hardware noise; it may also reflect the increasing difficulty of jointly optimizing the noisy quantum circuit and classical prediction head. The performance of the noisy models could also potentially be improved through error mitigation strategies \cite{prl-temme-2017-qem-short, prx-li-207-qem, revmod-cai-2023-qem}. We therefore leave a combined systematic study of the encoding design, depth-dependent optimizer tuning, and error mitigation as an important direction for future work.

\subsection{Classical Baselines}
Having established that the seasonal quantum encodings substantially outperform the standard encodings as reuploading depth increases and this robustness persists under noise, we finally compare the resulting quantum models with conventional classical forecasting approaches. We consider three complementary baselines: a random forest (RF), representing a nonlinear machine learning model; an autoregressive integrated moving average model, ARIMA(7,0,6), representing a statistical time series model; and a stacked long short-term memory (LSTM), representing a recurrent neural network with a comparable notion of depth. All classical models are trained and evaluated over the same chronological training, validation, and test data splits used on the same chronological training, validation, and test data splits as the quantum models. The implementation details are presented in the Supplemental Material.

\begin{figure*}
    \centering
    \includegraphics{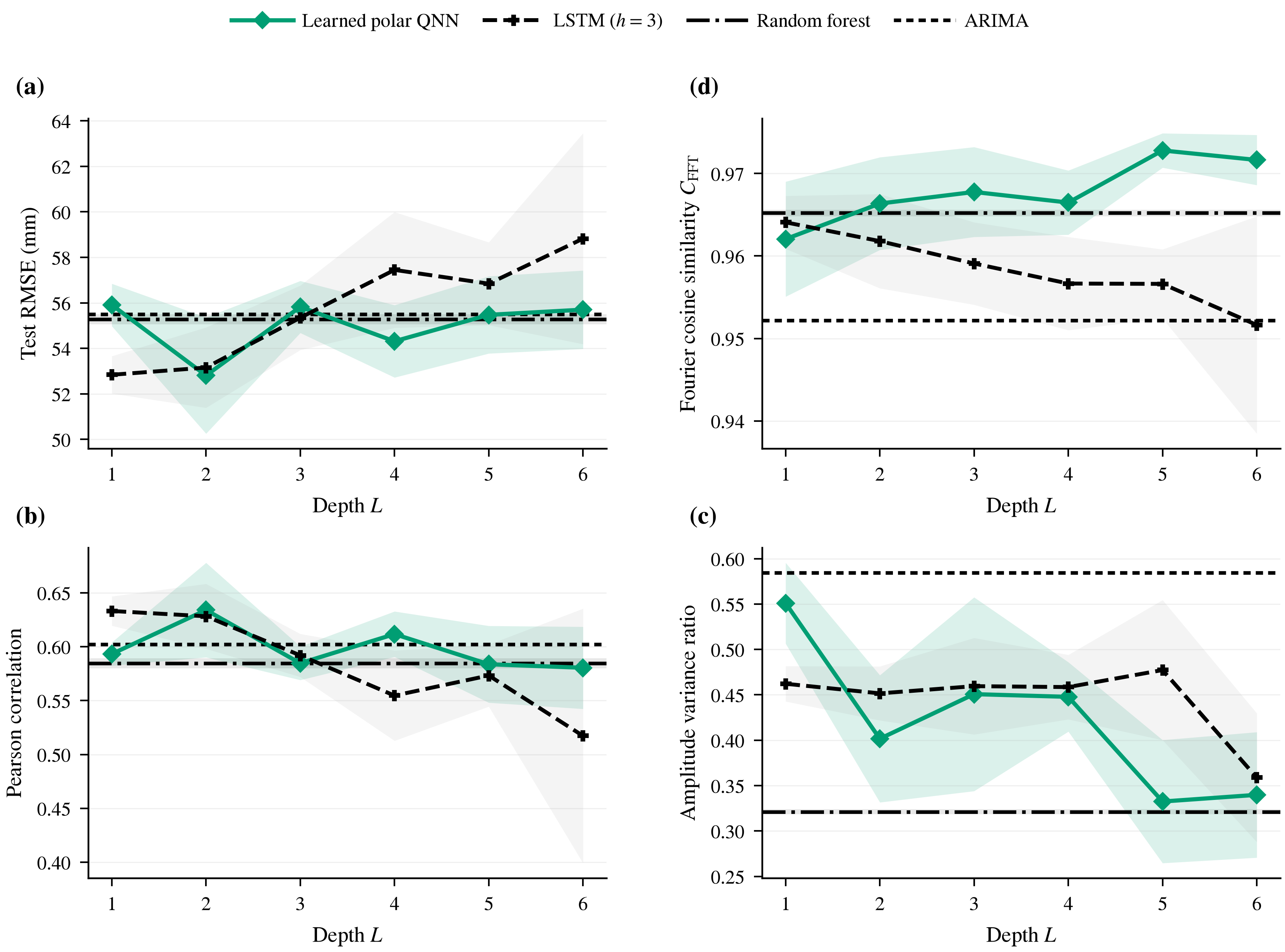}
    \caption{Comparison of the Learned Polar quantum data reuploading model with classical forecasting baselines across depth $L$. For the quantum model, $L$ denotes the number of repeated data-encoding and variational-circuit layers; for the LSTM, it denotes the number of stacked recurrent layers. The learned polar quantum model and LSTM are shown as depth-dependent trajectories, whereas random forest and ARIMA serve as depth-independent reference baselines. Shaded regions indicate \(\pm 1\) standard deviation across five independent realizations for the learned polar quantum model, LSTM, and ARIMA.}
    \label{fig:quantum, arima, lstm, rf}
\end{figure*}

The RF and ARIMA models provide depth-independent reference points for the forecasting task. For a more controlled comparison with the depth-dependent quantum models, we constructed stacked LSTMs with recurrent depth $L=1,\dots,6$ and hidden dimension $h=3$. The LSTM architecture was intentionally not chosen as a high-capacity, state-of-the-art forecasting model. Instead, its parameters were chosen so that it could serve as a classically comparable model to our learned quantum models, with a parameter count that grows approximately in step with that of the quantum model with equivalent reuploading depth.

Compared with these classical baseline models, the learned polar seasonal-encoding quantum model occupies the same overall forecasting regime, as shown in Fig.~\ref{fig:quantum, arima, lstm, rf}. In ideal simulations, the learnable polar achieved best test RMSEs of $52.81\pm2.55$. For comparison, the RF achieved $55.28\pm0.22~\mathrm{mm}$, ARIMA(7,0,6) achieved $55.50~\mathrm{mm}$, and the best mean LSTM result was $52.84\pm0.82~\mathrm{mm}$. The same overall picture is reflected in the correlation amplitude and spectral metrics: no classical baseline uniformly dominates across all forecast metrics, and the learned polar quantum model achieves the same performance. In the presence of noise, the performance of quantum model decreased, but still achieved an RMSE of $56.54~\mathrm{mm}$ and retained strong temporal and spectral structure. Together, these results show that the robustness introduced by seasonal quantum encoding is not at the expense of forecasting accuracy; rather, the resulting models remain competitive with established statistical and machine learning methods.

\section{Discussion}

In this work, we have demonstrated that the choice of encoding classical data into a quantum circuit can play a central role in determining how a data reuploading model behaves as circuit depth increases. For the rainfall time series considered here, standard value-based encodings progressively lose both seasonal structure and forecasting performance as additional reuploading layers are introduced. Explicitly incorporating the known seasonal recurrence pattern into the quantum feature map substantially changes this behavior. The fixed seasonal encoding shows improved performance at shallow depth, while the adaptive seasonal encodings remain far more robust across the full depth range. All adaptive seasonal encoding showed modest performance improvents at a depth of $L=2$ before declining at larger depths.  These results suggest that increasing circuit depth alone is not sufficient to improve a quantum time series model (in many cases it is detrimental to performance); the geometry through which temporally structured data are repeatedly presented to the circuit is equally (or more) important.

The geometric analysis provides a framework for interpreting this behavior. Two input windows separated by one seasonal period differ by a small displacement that is already present in the underlying time series, which prevents the time series from being perfectly periodic. The quantum circuit does not create this displacement but determines how strongly it is resolved by the input-space geometry induced by the model. Locally, this relation is captured by the Fubini-Study metric generated by the complete data reuploading circuit via a quantum geometric fidelity analysis. Since the effective input generators receive contributions from every reuploading layer, increasing depth can alter how the same underlying temporal differences are represented. However, preservation of seasonal fidelity should not be interpreted as the forecasting objective itself. Rather, the useful representation is one that preserves alignment of the quantumly encoded data with known seasonal structure. In this sense, the fidelity analysis provides a diagnostic of how temporal relationships are represented, rather than a direct measure of prediction quality.

These observations suggest a broader design principle for quantum learning of structured time series. When a physically or statistically meaningful temporal organization is already known, it can be advantageous to incorporate that structure directly into the quantum data encoding rather than requiring the variational circuit to infer it solely through repeated value-based encodings. In the present construction, precipitation magnitude determines the polar coordinate of the encoded qubit state while seasonal phase or calendar month determines the azimuthal coordinate, providing an explicit geometric representation of the annual cycle. The adaptive variants then deform this structured reference geometry during training. The fact that learning the polar deformation provides most of the observed improvement, while additional azimuthal flexibility gives no consistent advantage, further suggests that useful learning need not require unconstrained modification of every geometric degree of freedom. More generally, analogous constructions could incorporate other known temporal periods, phase variables, spatial coordinates, or physical symmetries into quantum representations.

The noisy simulations further show that the benefit of structured encoding is not limited to ideal state-vector evolution. At shallow and intermediate reuploading depths, the ordering established in the ideal simulations is retained under the IBM FakeAachen noise model, with the adaptive seasonal encodings continuing to outperform the fixed seasonal representation. At larger depths, however, the performance of all three seasonal models deteriorates, and their differences become increasingly compressed. This behavior highlights a practical trade-off among representational depth, hardware noise, and the increasing difficulty of hybrid SPSA optimization in a noisy landscape. A systematic separation of these effects, together with noise-aware optimization and error mitigation strategies, remains an important direction for future work.

In the presence of noise, the quantum performance decreases but remains close to the classical reference models at the best shallow and intermediate depths. These results demonstrate that the improved depth robustness achieved through structured quantum encoding can coexist with forecasting performance comparable to that of conventional classical approaches.

Finally, the classical benchmarks place the quantum results in the context of established forecasting methods. Random forests, ARIMA, and LSTM represent complementary standard statistical and machine learning approaches to the same prediction task. The best ideal seasonal quantum models achieve forecasting errors in family with these classical baselines. While it remains an open question whether QML methods can exhibit any quantum advantage, it is generally expected that quantum methods might be superior in analyzing correlated data sets \cite{schuld_is_2022}. The present comparison of the quantum algorithm with classical methods gives reason to be cautiously optimistic about the realizability of more intricate quantum machine learning protocols. Thus, an important next step is to extend our paradigm beyond univariate forecasting to multivariate time series, where temporal structure coexists with intervariable dependencies \cite{nonlineardyn-Sharma-2026-qml-multivar, supercomp-Ranilla-2026-qml-multivar}. In this setting, structure-aware quantum encodings could be used to investigate how temporal and cross-variable relationships are jointly represented, and how entangling operations provide useful representational structure \cite{nature-Laskar-2025-timeseries-ibmq, nature-Panadero-2024-express-qdr, nature-Kutvonen-2020-optimize-qreservoir, prxq-kobayashi-2024-feedback-qreservoir}. 

A complementary direction is to evaluate these models on real quantum hardware \cite{prl-hou-2026-exp-qreservoir, qmi-Ahmed-2025-optshot-qml,nature-Kandala-2017-hardware-vqc-magnets, arxiv-ibm-shallow-circuit_2026, arxiv-ionq-2026-clinical-qml,arxiv-quera-2026-analog-qml} and to develop more closely matched classical baselines that enable controlled comparisons with the corresponding quantum architectures \cite{arxiv-knitter-2026-classical-baseline}. Together, these extensions would test how broadly the encoding principles identified here apply as quantum time-series models move toward richer data and experimentally realized circuits.

\vspace{2em}

\section*{Data Availability}

The code for numerical simulations is available can be found at \cite{github_qml_dev_tripathy}.

\acknowledgements{S.D. acknowledges support from the John Templeton Foundation under Grant No. 63626.
During the course of the project, OpenAI's ChatGPT models were used to help understand Pennylane with PyTorch and Qiskit, to optimize code for HPCs and to orchestrate the project workflow. They were also used to help writing the plotting scripts for the figures.}

\bibliography{references}

\clearpage
\appendix

\onecolumngrid
%%%%%%%%%%%%%%%%%%%%%%%%%%%%%%%%%%%%%%%%%%%%%%%%%%%%%%
\section{Derivation of the local fidelity geometry}
\label{app:fidelity_geometry}
%%%%%%%%%%%%%%%%%%%%%%%%%%%%%%%%%%%%%%%%%%%%%%%%%%%%%%

Here, we present the details of the derivation for Eq.~\eqref{eq:fidelity_variance} in the main text: the relation between the fidelity of two approximately recurrent input windows and the effective generators of a depth $L$ data reuploading circuit. Consider two lag windows separated by the characteristic period $T_0$ of the time series,
\begin{equation}
    \mathbf{x}_{t+T_0}
    =
    \mathbf{x}_t
    +
    \boldsymbol{\delta}_t,
    \label{eq:app_seasonal_displacement}
\end{equation}
where $\boldsymbol{\delta}_t
    =
    \left(
        \delta_{t,1},
        \ldots,
        \delta_{t,w}
    \right)^T$ quantifies the displacement between the two windows.
Throughout the derivation, all derivatives are taken with respect to the continuous input coordinates entering the encoding operation, while the trained circuit parameters are held fixed. For compactness, we write $\ket{\psi}\equiv\ket{\psi_L(\mathbf{x}_t)}$ and evaluate all derivatives at $\mathbf{x}=\mathbf{x}_t$ unless
otherwise stated. For sufficiently small
$\boldsymbol{\delta}_t$, the state associated with the displaced
window has the Taylor expansion
\begin{equation}
    \ket{\psi_L(\mathbf{x}_t+\boldsymbol{\delta}_t)}
    =
    \ket{\psi}
    +
    \sum_{j=1}^{w}
    \delta_{t,j}
    \ket{\partial_j\psi}
    \nonumber
     +
    \frac{1}{2}
    \sum_{j,k=1}^{w}
    \delta_{t,j}\delta_{t,k}
    \ket{\partial_j\partial_k\psi}
    +
    \mathcal{O}
    \!\left(
        \|\boldsymbol{\delta}_t\|^3
    \right),
    \label{eq:app_state_expansion}
\end{equation}
where $\partial_j\equiv\frac{\partial}{\partial x_j}$.

%%%%%%%%%%%%%%%%%%%%%%%%%%%%%%%%%%%%%%%%%%%%%%%%%%%%%%
\subsection{Local expansion of the fidelity}
\label{app:fidelity_expansion}
%%%%%%%%%%%%%%%%%%%%%%%%%%%%%%%%%%%%%%%%%%%%%%%%%%%%%%

The fidelity between the two approximately recurrent windows is
\begin{equation}
    F_L(t,t+T_0)
    =
    \left|
        \left\langle
        \psi_L(\mathbf{x}_t)
        \middle|
        \psi_L(\mathbf{x}_t+\boldsymbol{\delta}_t)
        \right\rangle
    \right|^2.
    \label{eq:app_fidelity_definition}
\end{equation}
The corresponding overlap
amplitude is
\begin{equation}
    \mathcal{A}
    \equiv
    \left\langle
        \psi
        \middle|
        \psi_L(\mathbf{x}_t+\boldsymbol{\delta}_t)
    \right\rangle
     =
    1 +
    \sum_j
    \delta_{t,j}
    \braket{\psi}{\partial_j\psi}
    \nonumber\\
    +
    \frac{1}{2}
    \sum_{j,k}
    \delta_{t,j}\delta_{t,k}
    \braket{\psi}{\partial_j\partial_k\psi}
    +
    \mathcal{O}
    \!\left(
        \|\boldsymbol{\delta}_t\|^3
    \right).
    \label{eq:app_overlap_expansion}
\end{equation}
For notational convenience, define $c_j\equiv\braket{\psi}{\partial_j\psi}$, $ d_{jk}\equiv\braket{\psi}{\partial_j\partial_k\psi}$. Then,
\begin{equation}
    \mathcal{A}
    =
    1
    +
    \sum_j \delta_{t,j}c_j
    +
    \frac{1}{2}
    \sum_{j,k}
    \delta_{t,j}\delta_{t,k}d_{jk}
    +
    \mathcal{O}
    \!\left(
        \|\boldsymbol{\delta}_t\|^3
    \right).
    \label{eq:app_overlap_compact}
\end{equation}
The fidelity is obtained from
$F_L=\mathcal{A}\mathcal{A}^{*}$. Keeping terms up to second
order gives
\begin{equation}
    F_L(t,t+T_0)
    =
    1
    +
    \sum_j
    \delta_{t,j}
    \left(
        c_j+c_j^{*}
    \right)
    +
    \sum_{j,k}
    \delta_{t,j}\delta_{t,k}
    \left[
        c_j c_k^{*}
        +
        \frac{1}{2}
        \left(
            d_{jk}+d_{jk}^{*}
        \right)
    \right]
     +
    \mathcal{O}
    \!\left(
        \|\boldsymbol{\delta}_t\|^3
    \right).
    \label{eq:app_fidelity_intermediate}
\end{equation}

Using $\braket{\psi}{\psi}=1$, we can write $\braket{\partial_j\psi}{\psi} + \braket{\psi}{\partial_j\psi} = 0
$ or equivalently $c_j+c_j^{*}=0$
. Thus, the fidelity contains no term linear in
$\boldsymbol{\delta}_t$. Differentiating the normalization condition twice yields
\begin{align}
    0
    &=
    \partial_j\partial_k
    \braket{\psi}{\psi}
    \nonumber\\
    &=
    \braket{\partial_j\partial_k\psi}{\psi}
    +
    \braket{\partial_j\psi}{\partial_k\psi}
    +
    \braket{\partial_k\psi}{\partial_j\psi}
    +
    \braket{\psi}{\partial_j\partial_k\psi}.
    \label{eq:app_second_norm_derivative}
\end{align}
Taking the real part gives
\begin{equation}
    \operatorname{Re}(d_{jk})
    =
    -
    \operatorname{Re}
    \braket{\partial_j\psi}{\partial_k\psi}.
    \label{eq:app_second_normalization_identity}
\end{equation}

Substituting these identities into
Eq.~\eqref{eq:app_fidelity_intermediate}, and using the symmetry of $\delta_{t,j}\delta_{t,k}$ under interchange of $j$ and $k$, gives
\begin{equation}
    F_L(t,t+T_0)
    =
    1
    -
    \sum_{j,k}
    \delta_{t,j}\delta_{t,k}
    \operatorname{Re}
    \Big[
        \braket{\partial_j\psi}{\partial_k\psi}
    -
        \braket{\partial_j\psi}{\psi}
        \braket{\psi}{\partial_k\psi}
    \Big]
    +
    \mathcal{O}
    \!\left(
        \|\boldsymbol{\delta}_t\|^3
    \right).
    \label{eq:app_fidelity_fs_form}
\end{equation}
We therefore identify the input-space Fubini-Study metric induced by
the reuploading depth-$L$ circuit as
\begin{equation}
    g_{jk}^{(L)}
    =
    \operatorname{Re}
    \left[
        \braket{\partial_j\psi}{\partial_k\psi}
        -
        \braket{\partial_j\psi}{\psi}
        \braket{\psi}{\partial_k\psi}
    \right].
    \label{eq:app_fs_metric_derivative}
\end{equation}
Then Eq.~\eqref{eq:app_fidelity_fs_form} becomes
\begin{equation}
    F_L(t,t+T_0)
    =
    1
    -
    \sum_{j,k=1}^{w}
    g_{jk}^{(L)}
    \delta_{t,j}\delta_{t,k}
    +
    \mathcal{O}
    \!\left(
        \|\boldsymbol{\delta}_t\|^3
    \right),
    \label{eq:app_fidelity_metric}
\end{equation}or equivalently,
\begin{equation}
    1-F_L(t,t+T_0)
    =
    \boldsymbol{\delta}_t^T
    g^{(L)}
    \boldsymbol{\delta}_t
    +
    \mathcal{O}
    \!\left(
        \|\boldsymbol{\delta}_t\|^3
    \right).
    \label{eq:app_fidelity_quadratic_form}
\end{equation}

Equation~\eqref{eq:app_fidelity_quadratic_form} separates the
cycle-to-cycle displacement of the data,
$\boldsymbol{\delta}_t$, from the local geometry induced by the
quantum model, $g^{(L)}$. We next express this metric directly in
terms of the generators of the data-reuploading circuit.

%%%%%%%%%%%%%%%%%%%%%%%%%%%%%%%%%%%
\subsection{Effective input generator of a reuploading circuit}
\label{app:effective_generator}
%%%%%%%%%%%%%%%%%%%%%%%%%%%%%%%%%%%

The depth-$L$ circuit is $U_L(\mathbf{x})=A_L(\mathbf{x})A_{L-1}(\mathbf{x})\cdots A_1(\mathbf{x})$ where each reuploading block is
    $A_\ell(\mathbf{x})
    =
    V_\ell
    S_\ell(\mathbf{x})$.
The dependence on the trained circuit parameters is suppressed here for notational clarity. The variational operation $V_\ell$ does not
depend explicitly on the input coordinates.
For each input coordinate $x_j$ and encoding layer $\ell$, define the local encoding generator
\begin{equation}
    G_{j,\ell}
    \equiv
    i
    S_\ell^\dagger(\mathbf{x})
    \partial_j S_\ell(\mathbf{x}).
    \label{eq:app_bare_generator}
\end{equation}

Since $S_\ell$ is unitary, $S_\ell^\dagger S_\ell = I$. Differentiating this identity gives
\begin{equation}
    (\partial_jS_\ell^\dagger)S_\ell
    +
    S_\ell^\dagger(\partial_jS_\ell)
    =
    0,
\end{equation}
which shows that $S_\ell^\dagger\partial_jS_\ell$ is anti-Hermitian. Hence, $G_{j,\ell}$ defined in Eq.~\eqref{eq:app_bare_generator} is Hermitian. Multiplying Eq.~\eqref{eq:app_bare_generator} by $S_\ell$ from the left gives $\partial_j S_\ell = -iS_\ell G_{j,\ell}$. Because $V_\ell$ is independent of $\mathbf{x}$,
\begin{equation}
    \partial_jA_\ell
    =V_\ell\partial_jS_\ell
    =-iV_\ell S_\ell G_{j,\ell}
    =-iA_\ell G_{j,\ell}.
    \label{eq:app_derivative_layer}
\end{equation}
The same input coordinate is present in every reuploading layer.
Therefore,
\begin{align}
    \partial_j U_L
    &=
    \sum_{\ell=1}^{L}
    A_L\cdots A_{\ell+1}
    \left(
        \partial_jA_\ell
    \right)
    A_{\ell-1}\cdots A_1.
    \label{eq:app_derivative_UL_product}
\end{align}
Substituting Eq.~\eqref{eq:app_derivative_layer} gives
\begin{equation}
    \partial_j U_L
    =
    -i
    \sum_{\ell=1}^{L}
    A_L\cdots A_{\ell+1}
    A_\ell
    G_{j,\ell}
    A_{\ell-1}\cdots A_1.
    \label{eq:app_derivative_UL_generator}
\end{equation}

We now define the effective generator of the complete depth-$L$
circuit,
\begin{equation}
    \mathcal{G}_j^{(L)}
    \equiv
    iU_L^\dagger\partial_jU_L.
    \label{eq:app_effective_generator_definition}
\end{equation}
To evaluate it, define the partial circuit $B_\ell=A_\ell A_{\ell-1}\cdots A_1$, $B_0=I$. Using $U_L^\dagger
    =
    A_1^\dagger
    A_2^\dagger
    \cdots
    A_L^\dagger$, the contribution from reuploading layer $\ell$ becomes
\begin{align}
    &iU_L^\dagger
    \left[
        -i
        A_L\cdots A_{\ell+1}
        A_\ell
        G_{j,\ell}
        A_{\ell-1}\cdots A_1
    \right]
    \nonumber\\
    &=
    A_1^\dagger\cdots A_L^\dagger
    A_L\cdots A_{\ell+1}
    A_\ell
    G_{j,\ell}
    A_{\ell-1}\cdots A_1
    \nonumber\\
    &=
    A_1^\dagger\cdots A_{\ell-1}^\dagger
    G_{j,\ell}
    A_{\ell-1}\cdots A_1
    \nonumber\\
    &=
    B_{\ell-1}^\dagger
    G_{j,\ell}
    B_{\ell-1}.
    \label{eq:app_generator_single_contribution}
\end{align}
Summing over all reuploading layers gives
\begin{equation}
    \mathcal{G}_j^{(L)}
    =
    \sum_{\ell=1}^{L}
    B_{\ell-1}^\dagger
    G_{j,\ell}
    B_{\ell-1}.
    \label{eq:app_effective_generator}
\end{equation}
Thus, each reupload contributes a generator associated with the same input coordinate, but the contribution from layer $\ell$ is conjugated
by all circuit operations that precede that layer.

Equation~\eqref{eq:app_effective_generator_definition} also implies

\begin{equation}
    \partial_jU_L
    =
    -iU_L\mathcal{G}_j^{(L)}.
    \label{eq:app_U_derivative_effective}
\end{equation}
The depth-$L$ state is $\ket{\psi_L(\mathbf{x})} = U_L(\mathbf{x})\ket{\mathbf{0}}$, and therefore,
\begin{equation}
    \ket{\partial_j\psi} = 
    (\partial_jU_L)\ket{\mathbf{0}} = 
    -i
    U_L
    \mathcal{G}_j^{(L)}
    \ket{\mathbf{0}}.
    \label{eq:app_state_derivative_generator}
\end{equation}

%%%%%%%%%%%%%%%%%%%%%%%%%%%%%%%%%%%%%%%%%%%%%%%%%%%%%%
\subsection{Fubini-Study metric as a generator covariance}
\label{app:metric_covariance}
%%%%%%%%%%%%%%%%%%%%%%%%%%%%%%%%%%%%%%%%%%%%%%%%%%%%%%

We can now substitute Eq.~\eqref{eq:app_state_derivative_generator} directly into the
Fubini-Study metric of Eq.~\eqref{eq:app_fs_metric_derivative}. First,
\begin{equation}
    \braket{\partial_j\psi}{\partial_k\psi}
    =
    \bra{\mathbf{0}}
    \mathcal{G}_j^{(L)}
    U_L^\dagger U_L
    \mathcal{G}_k^{(L)}
    \ket{\mathbf{0}}
    =
    \bra{\mathbf{0}}
    \mathcal{G}_j^{(L)}
    \mathcal{G}_k^{(L)}
    \ket{\mathbf{0}}.
    \label{eq:app_derivative_overlap_generator}
\end{equation}
Similarly,
$\braket{\psi}{\partial_j\psi}
    =
    -i
    \bra{\mathbf{0}}
    \mathcal{G}_j^{(L)}
    \ket{\mathbf{0}}$
and hence,
\begin{align}
    \braket{\partial_j\psi}{\psi}
    \braket{\psi}{\partial_k\psi}
    &=
    \left\langle
        \mathcal{G}_j^{(L)}
    \right\rangle_0
    \left\langle
        \mathcal{G}_k^{(L)}
    \right\rangle_0,
    \label{eq:app_product_expectations}
\end{align}
where $\langle O\rangle_0\equiv\bra{\mathbf{0}}O\ket{\mathbf{0}}$. It follows that,
\begin{equation}
    g_{jk}^{(L)}
    =
    \operatorname{Re}
    \left\langle
        \mathcal{G}_j^{(L)}
        \mathcal{G}_k^{(L)}
    \right\rangle_0
    -
    \left\langle
        \mathcal{G}_j^{(L)}
    \right\rangle_0
    \left\langle
        \mathcal{G}_k^{(L)}
    \right\rangle_0.
    \label{eq:app_metric_generator_real}
\end{equation}
Since the effective generators are Hermitian,
\begin{equation}
    \operatorname{Re}
    \left\langle
        \mathcal{G}_j^{(L)}
        \mathcal{G}_k^{(L)}
    \right\rangle_0
    =
    \frac{1}{2}
    \left\langle
        \left\{
            \mathcal{G}_j^{(L)},
            \mathcal{G}_k^{(L)}
        \right\}
    \right\rangle_0.
\end{equation}
Defining the centered generators
    $\Delta\mathcal{G}_j^{(L)}
    =
    \mathcal{G}_j^{(L)}
    -
    \left\langle
        \mathcal{G}_j^{(L)}
    \right\rangle_0$ 
the metric takes the symmetrized covariance form
\begin{equation}
    g_{jk}^{(L)}
    =
    \frac{1}{2}
    \left\langle
        \left\{
            \Delta\mathcal{G}_j^{(L)},
            \Delta\mathcal{G}_k^{(L)}
        \right\}
    \right\rangle_0.
    \label{eq:app_metric_covariance}
\end{equation}

Substituting Eq.~\eqref{eq:app_metric_covariance} into
Eq.~\eqref{eq:app_fidelity_quadratic_form} gives
\begin{equation}
    1-F_L(t,t+T_0)
    =
    \frac{1}{2}
    \sum_{j,k}
    \delta_{t,j}\delta_{t,k}
    \left\langle
        \left\{
            \Delta\mathcal{G}_j^{(L)},
            \Delta\mathcal{G}_k^{(L)}
        \right\}
    \right\rangle_0
   +
    \mathcal{O}
    \!\left(
        \|\boldsymbol{\delta}_t\|^3
    \right).
    \label{eq:app_fidelity_covariance_expanded}
\end{equation}
To expose the structure along the actual seasonal displacement,
define the directional bare generator at layer $\ell$,
\begin{equation}
    G_{\boldsymbol{\delta},\ell}
    \equiv
    \sum_{j=1}^{w}
    \delta_{t,j}G_{j,\ell}.
    \label{eq:app_directional_bare_generator}
\end{equation}

Similarly, define the effective generator along
$\boldsymbol{\delta}_t$,
\begin{align}
    \mathcal{G}_{\boldsymbol{\delta}}^{(L)}
    &\equiv
    \sum_{j=1}^{w}
    \delta_{t,j}\mathcal{G}_j^{(L)}.
    \label{eq:app_directional_effective_definition}
\end{align}
Using Eq.~\eqref{eq:app_effective_generator},
\begin{align}
    \mathcal{G}_{\boldsymbol{\delta}}^{(L)}
    &=
    \sum_j
    \delta_{t,j}
    \sum_{\ell=1}^{L}
    B_{\ell-1}^\dagger
    G_{j,\ell}
    B_{\ell-1}
    \nonumber\\
    &=
    \sum_{\ell=1}^{L}
    B_{\ell-1}^\dagger
    \left(
        \sum_j
        \delta_{t,j}G_{j,\ell}
    \right)
    B_{\ell-1}
    \nonumber\\
    &=
    \sum_{\ell=1}^{L}
    B_{\ell-1}^\dagger
    G_{\boldsymbol{\delta},\ell}
    B_{\ell-1}.
    \label{eq:app_directional_effective_generator}
\end{align}
Because $\boldsymbol{\delta}_t$ is real,
Eq.~\eqref{eq:app_fidelity_covariance_expanded} can now be contracted
directly:
\begin{equation}
    \sum_{j,k}
    \delta_{t,j}\delta_{t,k}
    g_{jk}^{(L)}
    =
    \frac{1}{2}
    \left\langle
        \left\{
            \Delta
            \mathcal{G}_{\boldsymbol{\delta}}^{(L)},
            \Delta
            \mathcal{G}_{\boldsymbol{\delta}}^{(L)}
        \right\}
    \right\rangle_0
    =
    \left\langle
        \left(
            \Delta
            \mathcal{G}_{\boldsymbol{\delta}}^{(L)}
        \right)^2
    \right\rangle_0
    =
    \operatorname{Var}_0
    \left(
        \mathcal{G}_{\boldsymbol{\delta}}^{(L)}
    \right).
    \label{eq:app_quadratic_variance}
\end{equation}

The leading deviation from unit fidelity is therefore
\begin{equation}
    1-F_L(t,t+T_0)
    =
    \operatorname{Var}_0
    \left[
        \sum_{\ell=1}^{L}
        B_{\ell-1}^{\dagger}
        G_{\boldsymbol{\delta},\ell}
        B_{\ell-1}
    \right]
    +
    \mathcal{O}
    \!\left(
        \|\boldsymbol{\delta}_t\|^3
    \right).
    \label{eq:app_fidelity_variance_final}
\end{equation}

\section{Quantum data encodings}
\label{app:encodings}

This Appendix gives the explicit parameterizations used for the five
quantum data encodings considered in Sec.~\ref{sec:encodings} of the main text. All encodings act
independently on the \(w=14\) components of the lag window before the variational block at each data-reuploading layer. The value-based encodings depend solely on precipitation magnitude, whereas the three
seasonal encodings additionally depend on the position of each
observation within the annual cycle.
For a lag window
\begin{equation}
\mathbf{x}_t
=
\left(
y_{t-w+1},
\ldots,
y_t
\right)^T,
\end{equation}
the observation encoded on qubit
\(q\in\{0,\ldots,w-1\}\) has source index
\begin{equation}
r_{tq}
=
t-w+1+q.
\label{eq:app-source-index}
\end{equation}

For the present data set, \(w=14\). The seasonal index associated with
the same observation is
\begin{equation}
m_{tq}
=
r_{tq}\bmod T_0,
\qquad
m_{tq}\in\{0,\ldots,T_0-1\},
\label{eq:app-season-index}
\end{equation}
with seasonal period \(T_0=12\). For the two value-based encodings, precipitation is linearly mapped from
the physical interval $y\in[0,350]~\mathrm{mm}$ to
\begin{equation}
\widetilde{y}
=
2\frac{y}{350}-1,
\qquad
\widetilde{y}\in[-1,1].
\label{eq:app-linear-scaling}
\end{equation}
The corresponding rotation angle is
$\vartheta(y)
=
\pi\widetilde{y}$. At reuploading layer \(\ell\), the complete encoding operation has the
product form
\begin{equation}
S_\ell
=
\bigotimes_{q=0}^{w-1}
S_{\ell,q},
\label{eq:app-product-encoding}
\end{equation}
where the single-qubit map \(S_{\ell,q}\) depends on the encoding
strategy described below.

\subsection{\texorpdfstring{\(R_y\)}{Ry} encoding}
\label{app:ry-encoding}

The simplest encoding uses the scaled precipitation value as a rotation angle about the \(y\) axis,
\begin{equation}
S_{q}^{(R_y)}
\left(
y_{r_{tq}}
\right)
=
R_y
\left[
\pi\widetilde{y}_{r_{tq}}
\right]
\label{eq:app-ry-single}
\end{equation}
The complete encoding layer is, therefore,
\begin{equation}
S_{\ell}^{(R_y)}
\left(
\mathbf{x}_t
\right)
=
\bigotimes_{q=0}^{w-1}
R_y
\left[
\pi\widetilde{y}_{r_{tq}}
\right]
\label{eq:app-ry-window}
\end{equation}

This encoding contains no trainable parameters and is identical at every reuploading layer $\ell$. Each observation is represented along a
single Bloch sphere meridian.

\subsection{Same-scalar \texorpdfstring{\(R_xR_yR_z\)}{RxRyRz} encoding}
\label{app:rxryrz-encoding}

The second value-based encoding applies the same scalar angle to rotations
about all three axes. For a single observation,
\begin{equation}
S_{q}^{(R_xR_yR_z)}
\left(
y_{r_{tq}}
\right)
=
R_z
\left[
\pi\widetilde{y}_{r_{tq}}
\right]
R_y
\left[
\pi\widetilde{y}_{r_{tq}}
\right]
R_x
\left[
\pi\widetilde{y}_{r_{tq}}
\right].
\label{eq:app-rxryrz-single}
\end{equation}
Thus,
\begin{equation}
S_{\ell}^{(R_xR_yR_z)}
\left(
\mathbf{x}_t
\right)
=
\bigotimes_{q=0}^{w-1}
S_{q}^{(R_xR_yR_z)}
\left(
y_{r_{tq}}
\right).
\label{eq:app-rxryrz-window}
\end{equation}

Although this map generates a more intricate Bloch sphere trajectory
than the \(R_y\) encoding, the same value controls all three rotations. Consequently, the information supplied by each observation contains no explicit seasonal coordinate.

\subsection{Fixed seasonal CDF encoding}
\label{app:fixed-seasonal-encoding}

Precipitation magnitude is first transformed using the empirical
cumulative distribution function constructed exclusively from the raw
training observations. The training reference contains the \(364\)
observations with indices \(r=0,\ldots,363\), and the right-continuous
empirical CDF is
\begin{equation}
\mathrm{CDF}(y)
=
\frac{1}{364}
\sum_{r=0}^{363}
\mathbb{I}
\left(
y_r\leq y
\right).
\label{eq:app-empirical-cdf}
\end{equation}
The empirical rank is mapped to the interval \([-1,1]\) as
\begin{equation}
\zeta(y)
=
2\mathrm{CDF}(y)-1,
\label{eq:app-cdf-z}
\end{equation}
and subsequently to the polar angle
\begin{equation}
\alpha^{\mathrm{CDF}}(y)
=
\arccos
\left[
\zeta(y)
\right].
\label{eq:app-cdf-polar}
\end{equation}
For the observation on qubit \(q\),
$\alpha_{tq}^{\mathrm{CDF}}
=
\alpha^{\mathrm{CDF}}
\left(
y_{r_{tq}}
\right)$.
The seasonal phase is assigned to the azimuthal coordinate,
\begin{equation}
\phi_{tq}
=
\frac{2\pi m_{tq}}{T_0}
=
\frac{2\pi m_{tq}}{12}.
\label{eq:app-fixed-azimuth}
\end{equation}
The resulting single-qubit encoding is
\begin{equation}
S_{q}^{(\mathrm{CDF})}
=
R_z
\left(
\phi_{tq}
\right)
R_y
\left(
\alpha_{tq}^{\mathrm{CDF}}
\right),
\label{eq:app-fixed-seasonal-single}
\end{equation}
and the complete encoding layer is
\begin{equation}
S_{\ell}^{(\mathrm{CDF})}
\left(
\mathbf{x}_t
\right)
=
\bigotimes_{q=0}^{w-1}
R_z
\left(
\phi_{tq}
\right)
R_y
\left(
\alpha_{tq}^{\mathrm{CDF}}
\right).
\label{eq:app-fixed-seasonal-window}
\end{equation}

This encoding is fixed and therefore does not depend explicitly on
\(\ell\). The corresponding single-qubit Bloch vector is
\begin{equation}
\mathbf{r}_{tq}
=
\left(
\sqrt{1-\zeta_{tq}^{2}}\cos\phi_{tq},
\sqrt{1-\zeta_{tq}^{2}}\sin\phi_{tq},\zeta_{tq} \right)^T,
\label{eq:app-seasonal-bloch-vector}
\end{equation}
Thus, empirical precipitation rank controls the polar coordinate,
whereas the location within the annual cycle or calendar month controls the azimuthal coordinate.

\subsection{Learned polar seasonal CDF encoding}
\label{app:learned-polar-encoding}

The fixed seasonal encoder applies the same encoding
at every reuploading layer. To allow the representation to adapt to both
seasonal phase and circuit depth, the learned polar encoding introduces
two trainable parameters for every month-layer pair
\((m,\ell)\),
\begin{equation}
a_{m\ell},
\qquad
c_{m\ell},
\qquad
m=0,\ldots,11,
\qquad
\ell=1,\ldots,L.
\end{equation}
The first parameter determines a bounded multiplicative strength,
\begin{equation}
s_{m\ell}
=
1+\tanh
\left(
a_{m\ell}
\right),
\qquad
0<s_{m\ell}<2,
\label{eq:app-learned-strength}
\end{equation}
while the second determines a bounded polar displacement,
\begin{equation}
b_{m\ell}
=
\frac{\pi}{2}
\tanh
\left(
c_{m\ell}
\right),
\qquad
\left|
b_{m\ell}
\right|
<
\frac{\pi}{2}.
\label{eq:app-learned-polar-shift}
\end{equation}
For the observation encoded on qubit \(q\), the resulting month- and
layer-dependent polar angle is
\begin{equation}
\alpha_{tq,\ell}^{(\mathrm{LP})}
=
\frac{\pi}{2}
+
b_{m_{tq}\ell}
+
s_{m_{tq}\ell}
\left[
\alpha_{tq}^{\mathrm{CDF}}
-
\frac{\pi}{2}
\right].
\label{eq:app-learned-polar-angle}
\end{equation}
The fixed seasonal azimuthal coordinate is retained
\begin{equation}
\phi_{tq}^{(\mathrm{LP})}
=
\phi_{tq}
=
\frac{2\pi m_{tq}}{12}.
\end{equation}
The single qubit encoding at layer \(\ell\) is, therefore,
\begin{equation}
S_{\ell,q}^{(\mathrm{LP})}
=
R_z
\left(
\phi_{tq}^{(LP)}
\right)
R_y
\left(
\alpha_{tq,\ell}^{(\mathrm{LP})}
\right),
\label{eq:app-learned-polar-single}
\end{equation}
and the complete layer is
\begin{equation}
S_{\ell}^{(\mathrm{LP})}
=
\bigotimes_{q=0}^{w-1}
S_{\ell,q}^{(\mathrm{LP})}.
\label{eq:app-learned-polar-window}
\end{equation}

All raw encoding parameters are initialized at zero, $a_{m\ell} = c_{m\ell} =0$. Consequently, $s_{m\ell}=1, b_{m\ell}=0$, and Eq.~\eqref{eq:app-learned-polar-angle} reduces exactly to
\begin{equation}
\alpha_{tq,\ell}^{(\mathrm{LP})}
=
\alpha_{tq}^{\mathrm{CDF}}.
\end{equation}
The learned-polar model therefore begins from the fixed seasonal CDF
geometry and subsequently learns month- and layer-dependent
deformations of its polar coordinate.

\subsection{Learned polar $+$ azimuthal seasonal CDF encoding}
\label{app:learned-polar-azimuth-encoding}

The final encoding extends the learned-polar model by also allowing the
seasonal azimuthal coordinate to adapt during training. The polar angle is identical
to Eq.~\eqref{eq:app-learned-polar-angle}. In addition, each month-layer
pair is assigned a third raw trainable parameter $d_{m\ell}$ which determines a bounded azimuthal displacement,
\begin{equation}
\Delta\phi_{m\ell}
=
\frac{\pi}{12}
\tanh
\left(
d_{m\ell}
\right),
\qquad
\left|
\Delta\phi_{m\ell}
\right|
<
\frac{\pi}{12}.
\label{eq:app-bounded-azimuth-shift}
\end{equation}
The month- and layer-dependent azimuthal coordinate is
\begin{equation}
\phi_{tq,\ell}^{(\mathrm{LPA})}
=
\frac{2\pi m_{tq}}{12}
+
\Delta\phi_{m_{tq}\ell}.
\label{eq:app-learned-azimuth}
\end{equation}
The corresponding single-qubit encoding is
\begin{equation}
S_{\ell,q}^{(\mathrm{LPA})}
=
R_z
\left(
\phi_{tq,\ell}^{(\mathrm{LPA})}
\right)
R_y
\left(
\alpha_{tq,\ell}^{(\mathrm{LP})}
\right),
\label{eq:app-learned-polar-azimuth-single}
\end{equation}
and
\begin{equation}
S_{\ell}^{(\mathrm{LPA})}
=
\bigotimes_{q=0}^{w-1}
S_{\ell,q}^{(\mathrm{LPA})}.
\label{eq:app-learned-polar-azimuth-window}
\end{equation}

Adjacent fixed seasonal sectors are separated by $\pi/6$. Since the learned displacement satisfies $|\Delta\phi_{m\ell}|<\pi/12$, its maximum magnitude is half the angular separation between
adjacent month sectors. The azimuthal coordinate can therefore adapt locally while
remaining bounded around the corresponding fixed seasonal direction.

All three raw encoding parameters are initialized at zero,
$a_{m\ell}
=
c_{m\ell}
=
d_{m\ell}
=
0$,
which gives $
s_{m\ell}=1,
\qquad
b_{m\ell}=0,
\qquad
\Delta\phi_{m\ell}=0.
$. Thus, the learned polar $+$ azimuthal model also reduces exactly to the
fixed seasonal CDF encoding at initialization.

\section{Noisy Simulation Details}
\label{sec:noisy-implementation}

The finite-shot noisy simulations are carried out with the IBM FakeAachen through the PennyLane-Qiskit interface. Each expectation value is estimated from \(S=512\) measurement shots.

All circuits are transpiled to the native FakeAachen basis using transpiler optimization level~\(1\) and SABRE routing. This implementation uses the fixed physical qubit layout \[
[33,32,31,18,11,10,9,8,7,17,27,28,29,38].
\]
The closing logical ring CNOT interaction is therefore routed through the available
physical connectivity whenever the corresponding physical qubits are not
directly coupled. Simulator truncation is enabled so that inactive physical
qubits do not contribute to the simulated state space.

\subsection{Hybrid SPSA-Adam optimization}
\label{sec:spsa-adam}

Let \(\bm{\theta}\in\mathbb{R}^{P_{\mathrm q}}\) denote the complete set of trainable quantum parameters. This vector contains the variational circuit
parameters and, for the adaptive encodings, the trainable encoding parameters. We
denote the parameters of the classical prediction head by
$
\bm{\phi} = \left(
w_{\mathrm h}, b_{\mathrm h}\right)$. The quantum parameters are optimized using simultaneous perturbation
stochastic approximation (SPSA), while the classical prediction head is
optimized using Adam. At SPSA iteration \(k\), the learning rate and
perturbation schedules are
\begin{equation}
a_k
=
\frac{a}{(A+k)^{\alpha}},
\qquad
c_k
=
\frac{c}{k^{\gamma}},
\label{eq:spsa-schedules}
\end{equation}
with
\begin{equation}
a=0.40,
\qquad
c=0.05,
\qquad
A=50,
\qquad
\alpha=0.602,
\qquad
\gamma=0.101.
\end{equation}
At each iteration, a perturbation vector
\(\bm{\Delta}_k\in\{-1,+1\}^{P_{\mathrm q}}\) is drawn with independent Rademacher components. Two perturbed quantum-parameter vectors are then
constructed,
\begin{equation}
\bm{\theta}_k^{\pm}
=
\bm{\theta}_k
\pm
c_k\bm{\Delta}_k.
\label{eq:spsa-perturbations}
\end{equation}
For a mini-batch \(\mathcal{B}_k\), the noisy quantum circuit is evaluated once
at each perturbed parameter vector. The corresponding losses are
\begin{equation}
\mathcal{L}_k^{\pm}
=
\frac{1}{|\mathcal{B}_k|}
\sum_{t\in\mathcal{B}_k}
\left[
h_{\bm{\phi}_k}
\!\left(
Q_{\bm{\theta}_k^{\pm}}(\mathbf{x}_t)
\right)
-
\widetilde{y}_{t+1}
\right]^2,
\label{eq:spsa-perturbed-losses}
\end{equation}
where \(Q_{\bm{\theta}}(\mathbf{x})\) denotes the finite-shot vector of quantum
features after measurement, \(h_{\bm{\phi}}\) is the classical prediction head, and
\(\widetilde{y}_{t+1}\) denotes the scaled/transformed target used during training.

The SPSA estimator of the quantum gradient is
\begin{equation}
\widehat{\bm{g}}_{\bm{\theta},k}
= \Bigg(
\frac{
\mathcal{L}_k^{+}-\mathcal{L}_k^{-}
}{
2c_k
} \Bigg) 
\bm{\Delta}_k^{-1},
\label{eq:spsa-gradient}
\end{equation}
where the inverse is componentwise. Because each component of
\(\bm{\Delta}_k\) is either \(+1\) or \(-1\),
\(\bm{\Delta}_k^{-1}=\bm{\Delta}_k\). The quantum parameters are then updated as
\begin{equation}
\bm{\theta}_{k+1}
=
\bm{\theta}_k
-
a_k\widehat{\bm{g}}_{\bm{\theta},k}.
\label{eq:spsa-update}
\end{equation}

The classical-head loss is defined as the symmetric average of the two
perturbed losses,
\begin{equation}
\mathcal{L}_{\mathrm{head},k}
=
\frac{1}{2}
\left(
\mathcal{L}_k^{+}
+
\mathcal{L}_k^{-}
\right).
\label{eq:head-symmetric-loss}
\end{equation}
The quantum features are detached from the PyTorch computational graph before
the classical backward pass, so automatic differentiation acts only on
\(\bm{\phi}\). The classical parameters are updated with Adam using learning
rate \(\eta_{\mathrm{Adam}}=0.005\), and the norm of the classical
gradient is clipped to \(1.0\) before each update.

Both the SPSA estimator and the Adam gradient are constructed from the same two
quantum evaluations at \(\bm{\theta}_k^{+}\) and \(\bm{\theta}_k^{-}\). The complete training procedure
is summarized below.

\begin{center}
\begin{minipage}{0.96\linewidth}
\hrule
\vspace{0.5em}
\textbf{Algorithm: Finite-shot hybrid SPSA-Adam training}
\vspace{0.25em}

\begin{algorithmic}[1]
\Require Training set
\(\mathcal{D}_{\mathrm{tr}}
=\{(\mathbf{x}_t,y_{t+1})\}\)
\Require Initial quantum parameters \(\bm{\theta}_1\)
\Require Initial classical head parameters \(\bm{\phi}_1\)
\Require SPSA constants \(a,c,A,\alpha,\gamma\)
\Require Adam learning rate \(\eta_{\mathrm{Adam}}\)

\For{\(e=1,\ldots,E\)}
    \State Shuffle the training windows

    \For{each mini-batch \(\mathcal{B}_k\)}
        \State \(a_k\gets a/(A+k)^{\alpha}\)
        \State \(c_k\gets c/k^{\gamma}\)

        \State Sample
        \(\bm{\Delta}_k\in\{-1,+1\}^{P_{\mathrm q}}\)

        \State
        \(\bm{\theta}_k^{+}
        \gets
        \bm{\theta}_k+c_k\bm{\Delta}_k\)

        \State
        \(\bm{\theta}_k^{-}
        \gets
        \bm{\theta}_k-c_k\bm{\Delta}_k\)

        \State Evaluate
        \(Q_{\bm{\theta}_k^{+}}(\bm{x}_t)\)
        for all \(t\in\mathcal{B}_k\)

        \State Evaluate
        \(Q_{\bm{\theta}_k^{-}}(\bm{x}_t)\)
        for all \(t\in\mathcal{B}_k\)

        \State Compute
        \(\mathcal{L}_k^{+}\) and \(\mathcal{L}_k^{-}\)
        using Eq.~\eqref{eq:spsa-perturbed-losses}

        \State
        \(\mathcal{L}_{\mathrm{head},k}
        \gets
        (\mathcal{L}_k^{+}+\mathcal{L}_k^{-})/2\)

        \State Detach the quantum features from the PyTorch graph

        \State Compute
        \(\nabla_{\bm{\phi}}
        \mathcal{L}_{\mathrm{head},k}\)
        by automatic differentiation

        \State Clip
        \(\lVert
        \nabla_{\bm{\phi}}
        \mathcal{L}_{\mathrm{head},k}
        \rVert_2\)
        to \(1.0\)

        \State Update
        \(\bm{\phi}_{k+1}\)
        from \(\bm{\phi}_k\) using Adam

        \State Compute
        \(\widehat{\bm{g}}_{\bm{\theta},k}\)
        using Eq.~\eqref{eq:spsa-gradient}

        \State
        \(\bm{\theta}_{k+1}
        \gets
        \bm{\theta}_k
        -
        a_k\widehat{\bm{g}}_{\bm{\theta},k}\)

        \State Increment \(k\)
    \EndFor

    \If{\(e=1\), \(e=E\), or \(e\) is a multiple of \(5\)}
        \State Evaluate the center-point validation loss
        \(\mathcal{L}_{\mathrm{val}}
        (\bm{\theta}_k,\bm{\phi}_k)\)

        \If{the validation loss has improved}
            \State Save the current model as the best checkpoint
        \EndIf
    \EndIf

    \State Save the resumable training checkpoint
\EndFor

\State Restore the checkpoint with the lowest validation loss
\State Evaluate the restored model once on the test set
\end{algorithmic}

\vspace{0.5em}
\hrule
\end{minipage}
\end{center}

Finite-shot quantum evaluations are distributed across
\(N_{\mathrm{worker}}=32\) worker processes. Each worker constructs
and retains its own PennyLane QNode and Qiskit Aer backend. For a mini batch of \(16\) training windows, the positive and negative SPSA perturbations require $2|\mathcal{B}_k|= 32$
quantum circuit evaluations. These evaluations are submitted
concurrently to the \(32\) CPU workers. The transpiler seed is identical across workers, ensuring that all workers use
the same compiled logical circuit. In contrast, the simulator seed assigned to
worker \(j\) is \(s_j =
s_{\mathrm{base}}+j, \qquad j=0,\ldots, N_{\mathrm{worker}}-1\), which provides independent finite-shot and noise realizations across workers. With \(350\) training windows and mini batch size \(16\), each epoch contains $ N_{\mathrm{step}}
=
\left\lceil
\frac{350}{16}
\right\rceil
=
22$ SPSA updates. Because each update evaluates two perturbed quantum circuits, training
requires \(700\) quantum evaluations per epoch and \(70\,000\)
training evaluations over \(100\) epochs for each independent run.

\newcommand{\NoisyColA}[1]{%
\parbox[t]{0.18\textwidth}{\raggedright #1}}
\newcommand{\NoisyColB}[1]{%
\parbox[t]{0.24\textwidth}{\raggedright #1}}
\newcommand{\NoisyColC}[1]{%
\parbox[t]{0.50\textwidth}{\raggedright #1}}

\begin{longtable}{@{}lll@{}}
\caption{ Noisy Simulation Details
}
\label{tab:noisy-parameters}\\
\toprule
\NoisyColA{Category}
&
\NoisyColB{Parameter}
&
\NoisyColC{Value or specification}
\\
\midrule
\endfirsthead

\multicolumn{3}{c}{
\tablename\ \thetable\ --- continued from previous page
}
\\
\toprule
\NoisyColA{Category}
&
\NoisyColB{Parameter}
&
\NoisyColC{Value or specification}
\\
\midrule
\endhead

\midrule
\multicolumn{3}{r}{Continued on next page}
\\
\endfoot

\bottomrule
\endlastfoot

\NoisyColA{Data}
&
\NoisyColB{Time-series length}
&
\NoisyColC{\(539\) monthly observations}
\\

\NoisyColA{Data}
&
\NoisyColB{Lag-window length}
&
\NoisyColC{\(w=14\)}
\\

\NoisyColA{Data}
&
\NoisyColB{Training windows}
&
\NoisyColC{\(350\)}
\\

\NoisyColA{Data}
&
\NoisyColB{Validation windows}
&
\NoisyColC{\(70\)}
\\

\NoisyColA{Data}
&
\NoisyColB{Test windows}
&
\NoisyColC{\(105\)}
\\

\NoisyColA{Data}
&
\NoisyColB{Target range used for scaling}
&
\NoisyColC{\([0,350]\,\mathrm{mm}\)}
\\

\NoisyColA{Data}
&
\NoisyColB{Scaled target range}
&
\NoisyColC{\([-1,1]\)}
\\

\NoisyColA{Data}
&
\NoisyColB{Forecast horizon}
&
\NoisyColC{One month}
\\

\midrule

\NoisyColA{Quantum model}
&
\NoisyColB{Number of logical qubits}
&
\NoisyColC{\(14\)}
\\

\NoisyColA{Quantum model}
&
\NoisyColB{Reuploading depths}
&
\NoisyColC{\(L\in\{1,\ldots,6\}\)}
\\

\NoisyColA{Quantum model}
&
\NoisyColB{Ansatz layers per reupload}
&
\NoisyColC{\(1\)}
\\

\NoisyColA{Quantum model}
&
\NoisyColB{Single-qubit ansatz}
&
\NoisyColC{
\(\operatorname{Rot}
(\theta^{(1)},\theta^{(2)},\theta^{(3)})\)
on every qubit
}
\\

\NoisyColA{Quantum model}
&
\NoisyColB{Entangling pattern}
&
\NoisyColC{
Nearest-neighbor CNOT chain with closing interaction
\(\operatorname{CNOT}_{13,0}\)
}
\\

\NoisyColA{Quantum model}
&
\NoisyColB{Measurement}
&
\NoisyColC{
Individual Pauli-\(Z\) expectation values on all \(14\) qubits
}
\\

\NoisyColA{Quantum model}
&
\NoisyColB{Classical head}
&
\NoisyColC{
\(\operatorname{Linear}(14,1)\) followed by \(\tanh\)
}
\\

\NoisyColA{Quantum model}
&
\NoisyColB{Variational initialization}
&
\NoisyColC{
Independent zero-centered initialization with scale \(0.01\)
}
\\

\NoisyColA{Quantum model}
&
\NoisyColB{Learned-encoding initialization}
&
\NoisyColC{
Raw strength, polar-shift, and azimuth-shift parameters
initialized to zero
}
\\

\NoisyColA{Quantum model}
&
\NoisyColB{Quantum parameters}
&
\NoisyColC{
\(42L\) for fixed encoding,
\(66L\) for learned polar, and
\(78L\) for learned polar plus azimuth
}
\\

\NoisyColA{Quantum model}
&
\NoisyColB{Classical parameters}
&
\NoisyColC{\(15\)}
\\

\midrule

\NoisyColA{Noisy simulation}
&
\NoisyColB{Backend mode}
&
\NoisyColC{Qiskit Aer finite-shot noisy simulation}
\\

\NoisyColA{Noisy simulation}
&
\NoisyColB{Noise source}
&
\NoisyColC{IBM FakeAachen}
\\

\NoisyColA{Noisy simulation}
&
\NoisyColB{Physical backend size}
&
\NoisyColC{\(156\) qubits}
\\

\NoisyColA{Noisy simulation}
&
\NoisyColB{Physical layout}
&
\NoisyColC{
\([33,32,31,18,11,10,9,8,7,17,27,28,29,38]\)
}
\\

\NoisyColA{Noisy simulation}
&
\NoisyColB{Routing method}
&
\NoisyColC{SABRE}
\\

\NoisyColA{Noise simulation}
&
\NoisyColB{Transpiler optimization level}
&
\NoisyColC{\(1\)}
\\

\NoisyColA{Noisy simulation}
&
\NoisyColB{Shots}
&
\NoisyColC{\(512\) per expectation-value estimate}
\\

\NoisyColA{Noise simulation}
&
\NoisyColB{Aer method}
&
\NoisyColC{\texttt{automatic}}
\\

\NoisyColA{Noisy simulation}
&
\NoisyColB{Simulator truncation}
&
\NoisyColC{Enabled}
\\

\midrule

\NoisyColA{Training}
&
\NoisyColB{Independent seeds}
&
\NoisyColC{\(42\)}
\\

\NoisyColA{Training}
&
\NoisyColB{Epochs}
&
\NoisyColC{\(100\)}
\\

\NoisyColA{Training}
&
\NoisyColB{Mini-batch size}
&
\NoisyColC{\(16\)}
\\

\NoisyColA{Training}
&
\NoisyColB{Updates per epoch}
&
\NoisyColC{\(22\)}
\\

\NoisyColA{Training}
&
\NoisyColB{SPSA updates per run}
&
\NoisyColC{\(2200\)}
\\

\NoisyColA{Training}
&
\NoisyColB{Loss}
&
\NoisyColC{Mean-squared error}
\\

\NoisyColA{Training}
&
\NoisyColB{Quantum optimizer}
&
\NoisyColC{SPSA}
\\

\NoisyColA{Training}
&
\NoisyColB{SPSA coefficient \(a\)}
&
\NoisyColC{\(0.40\)}
\\

\NoisyColA{Training}
&
\NoisyColB{SPSA perturbation coefficient \(c\)}
&
\NoisyColC{\(0.05\)}
\\

\NoisyColA{Training}
&
\NoisyColB{SPSA stability constant \(A\)}
&
\NoisyColC{\(50\)}
\\

\NoisyColA{Training}
&
\NoisyColB{SPSA decay exponent \(\alpha\)}
&
\NoisyColC{\(0.602\)}
\\

\NoisyColA{Training}
&
\NoisyColB{SPSA perturbation exponent \(\gamma\)}
&
\NoisyColC{\(0.101\)}
\\

\NoisyColA{Training}
&
\NoisyColB{Classical optimizer}
&
\NoisyColC{Adam}
\\

\NoisyColA{Training}
&
\NoisyColB{Adam learning rate}
&
\NoisyColC{\(0.005\)}
\\

\NoisyColA{Training}
&
\NoisyColB{Classical gradient clipping}
&
\NoisyColC{Maximum Euclidean norm \(1.0\)}
\\

\NoisyColA{Training}
&
\NoisyColB{Validation schedule}
&
\NoisyColC{Epoch \(1\), every fifth epoch, and the final epoch}
\\

\NoisyColA{Training}
&
\NoisyColB{Validation repeats}
&
\NoisyColC{\(1\)}
\\

\NoisyColA{Training}
&
\NoisyColB{Validation batch size}
&
\NoisyColC{\(32\)}
\\

\NoisyColA{Training}
&
\NoisyColB{Checkpoint interval}
&
\NoisyColC{Every epoch}
\\

\NoisyColA{Training}
&
\NoisyColB{Model selection}
&
\NoisyColC{Lowest validation mean-squared error}
\\

\end{longtable}

\let\NoisyColA\relax
\let\NoisyColB\relax
\let\NoisyColC\relax

\section{Forecasting metrics}
\label{app:forecasting-metrics}

Forecasting performance was evaluated on the test set using four complementary metrics. Let \(y_t\) and \(\widehat{y}_t\) denote the observed and predicted precipitation, respectively, for \(t=0,\ldots,N_{\mathrm{test}}-1\), where \(N_{\mathrm{test}}=105\).

The test root-mean-square error is defined as
\begin{equation}
\text{ Test RMSE}
=
\sqrt{
\frac{1}{N_{\mathrm{test}}}
\sum_{t=0}^{N_{\mathrm{test}}-1}
\left(
\widehat{y}_t-y_t
\right)^2
}.
\label{eq:test-rmse}
\end{equation}
RMSE was computed after restoring model predictions to the original precipitation scale and is therefore reported in millimeters.

To quantify similarity in frequency domain structure, the observed and predicted test sequences were first centered independently by subtracting their respective means,
\begin{equation}
y_t^{\,c}=y_t-\overline{y},
\qquad
\widehat{y}_t^{\,c}
=
\widehat{y}_t-\overline{\widehat{y}}.
\end{equation}
Their discrete Fourier transforms were computed as
\begin{align}
\widetilde{y}_k
&=
\sum_{t=0}^{N_{\mathrm{test}}-1}
y_t^{\,c}
\exp\left(
-\frac{2\pi i kt}{N_{\mathrm{test}}}
\right),
\\
\widetilde{\widehat{y}}_k
&=
\sum_{t=0}^{N_{\mathrm{test}}-1}
\widehat{y}_t^{\,c}
\exp\left(
-\frac{2\pi i kt}{N_{\mathrm{test}}}
\right).
\end{align} 
The corresponding Fourier power spectra are
\begin{equation}
P_k
=
\left|\widetilde{y}_k\right|^2,
\qquad
\widehat{P}_k
=
\left|\widetilde{\widehat{y}}_k\right|^2.
\end{equation}
The Fourier cosine similarity was evaluated over the non-zero positive frequency bins \(\mathcal{K}_{+}\),
\begin{equation}
C_{\mathrm{FFT}}
=
\frac{
\displaystyle
\sum_{k\in\mathcal{K}_{+}}
P_k\widehat{P}_k
}{
\displaystyle
\sqrt{
\sum_{k\in\mathcal{K}_{+}}P_k^2
}
\sqrt{
\sum_{k\in\mathcal{K}_{+}}\widehat{P}_k^2
}
}.
\label{eq:fft-cosine}
\end{equation}

Since the Fourier power is nonnegative,
\(0\leq C_{\mathrm{FFT}}\leq1\), with
\(C_{\mathrm{FFT}}=1\) indicating identical normalized spectral shapes. This quantity therefore measures whether the forecast reproduces the relative distribution of power across temporal frequencies. Because it is a cosine similarity, it is insensitive to an overall multiplicative rescaling of the predicted power spectrum. This means that high spectral similarity does not by itself imply agreement in forecast amplitude.

Temporal alignment between the observed and predicted sequences was quantified using the Pearson correlation coefficient,
\begin{equation}
r_{\mathrm{P}}
=
\frac{
\displaystyle
\sum_{t=0}^{N_{\mathrm{test}}-1}
\left(y_t-\overline{y}\right)
\left(
\widehat{y}_t-\overline{\widehat{y}}
\right)
}{
\displaystyle
\sqrt{
\sum_{t=0}^{N_{\mathrm{test}}-1}
\left(y_t-\overline{y}\right)^2
}
\sqrt{
\sum_{t=0}^{N_{\mathrm{test}}-1}
\left(
\widehat{y}_t-\overline{\widehat{y}}
\right)^2
}
}.
\label{eq:pearson-correlation}
\end{equation}
The coefficient satisfies
\(-1\leq r_{\mathrm{P}}\leq1\), with
\(r_{\mathrm{P}}=1\) corresponding to perfect positive linear alignment between the observed and predicted temporal fluctuations. Because the means and overall scales of the two sequences are removed in the correlation calculation, \(r_{\mathrm{P}}\) primarily measures whether the forecast tracks the temporal variation of the observations rather than their absolute magnitude.

Finally, recovery of the observed forecast amplitude was quantified using the amplitude variance ratio,
\begin{equation}
R_{\mathrm{var}}
=
\frac{
\operatorname{Var}\!\left(\widehat{y}\right)
}{
\operatorname{Var}(y)
},
\label{eq:variance-ratio}
\end{equation}
where the variances were evaluated over the complete test interval using
$\operatorname{Var}(y)
=
\frac{1}{N_{\mathrm{test}}}
\sum_{t=0}^{N_{\mathrm{test}}-1}
\left(y_t-\overline{y}\right)^2$
and analogously for \(\widehat{y}\). The ideal value is \(R_{\mathrm{var}}=1\). Values below 1 indicate suppression of the observed variability and therefore forecasts with reduced amplitude, whereas values above unity indicate variability larger than that of the observed precipitation series.

These four metrics probe distinct aspects of forecast quality: Test RMSE measures absolute pointwise error, \(C_{\mathrm{FFT}}\) measures normalized frequency-domain structure, \(r_{\mathrm{P}}\) measures temporal alignment, and \(R_{\mathrm{var}}\) measures recovery of the observed variation in amplitude.

\section{Classical forecasting baselines}
\label{app:classical-baselines}

To compare the performance of our quantum models, we evaluated three complementary classical baselines: random forest (RF), autoregressive integrated moving average (ARIMA), and long short-term memory (LSTM). The classical models were evaluated over the same chronological training, validation, and test splits used for the quantum models. No test observations were used for hyperparameter or model selection. For the random forest and LSTM models, the precipitation series was represented using the same supervised lag window construction as the quantum models. Each input contained the preceding \(w=14\) monthly precipitation observations,
\begin{equation}
\mathbf{x}_t
=
\left(
y_{t-13},\ldots,y_t
\right)^T,
\end{equation}
with the subsequent observation \(y_{t+1}\) used as the one-month-ahead target. This resulted in \(350\) training, \(70\) validation, and \(105\) test forecasting examples.

Precipitation values were transformed to the interval \([-1,1]\) according to
\begin{equation}
\widetilde{y}
=
2\frac{y}{350}-1,
\label{eq:app-classical-scaling}
\end{equation}
where \(y\) is measured in millimeters. Predictions were transformed back to physical units using $y = 175\left(\widetilde{y}+1\right)$. Random forest and LSTM are stochastic models and were evaluated over five independent runs using random seeds \(42,43,44,45,\) and \(46\). The source of stochasticity is model-dependent: for the LSTM, the random state determines the initialization of the trainable parameters and the shuffled mini-batch sequence during training. For the random forest, it determines the stochastic bootstrap of the tree ensemble. The ARIMA model is deterministic for a fixed model specification and was consequently evaluated using a single fit. All classical models were evaluated using the same set of forecasting metrics used for the quantum models.

\subsection{Random forest}
\label{app:random-forest}

The random-forest baseline was implemented using the \texttt{RandomForestRegressor} in scikit-learn. Each \(14\)-month lag window was supplied to the model as a \(14\)-dimensional feature vector. Individual regression trees used the squared-error criterion and bootstrap sampling. Model hyperparameters were selected using a grid search over the validation set, with the parameter search including
\begin{align}
n_{\mathrm{estimators}}
&\in
\{100,300,500,1000\},
\\
d_{\max}
&\in
\{3,5,8,10,15,\mathrm{None}\},
\\
n_{\mathrm{leaf}}^{\min}
&\in
\{1,2,4,8,12,16\},
\\
f_{\mathrm{feat}}
&\in
\{0.25,0.50,0.75,1.0\},
\end{align}
where \(n_{\mathrm{estimators}}\) denotes the number of trees, \(d_{\max}\) the maximum tree depth, \(n_{\mathrm{leaf}}^{\min}\) the minimum number of samples in a terminal leaf, and \(f_{\mathrm{feat}}\) the fraction of input features available when determining a split. The complete grid therefore contained \(576\) hyperparameter configurations. Each configuration was trained independently for the five random states \(42,\ldots,46\). The selected configuration, according to the mean validation mean-squared error, was
\begin{equation}
n_{\mathrm{estimators}}=500,\qquad
d_{\max}=5,\qquad
n_{\mathrm{leaf}}^{\min}=2,\qquad
f_{\mathrm{feat}}=1.0.
\end{equation}

\subsection{LSTM}
\label{app:lstm}

Each \(14\) month input window was represented as a sequence of length \(14\) with one scalar feature per time step, so that the input to the recurrent network had shape $14\times1$. The input dimension and hidden dimension were fixed to $n_{\mathrm{in}}=1, h=3$, while the number of stacked recurrent layers was varied over $L=1,\ldots,6$. All LSTMs were unidirectional and used no dropout. After processing the complete \(14\) month sequence, the final hidden state was passed to a linear output layer, followed by a tanh activation to produce the forecast in the scaled target space. The models were trained by minimizing mean squared error using Adam with learning rate $\eta=0.005$. For every depth \(L\), five independent runs were performed using random states \(42,\ldots,46\). The best model was chosen using minimum validation loss during training.

\begin{table}[t]
\centering

\begin{tabular}{c c c c c}
\toprule
\textbf{Depth}\quad &
\quad\textbf{LSTM}\quad &
\quad\textbf{Fixed Seasonal}\quad &
\quad\textbf{Learned Polar}\quad &
\quad\textbf{Learned Polar $+$ Azimuthal} \\
\midrule
1 & 76  & 57  & 81  & 93  \\
2 & 172 & 99  & 147 & 171 \\
3 & 268 & 141 & 213 & 249 \\
4 & 364 & 183 & 279 & 327 \\
5 & 460 & 225 & 345 & 405 \\
6 & 556 & 267 & 411 & 483 \\
\bottomrule
\end{tabular}
\caption{
Total trainable parameter counts for the LSTM and the three seasonal encoding models across model depth.
The LSTM hidden dimension was fixed at $h=3$. LSTM depth and quantum reuploading depth are not architecturally equivalent and are aligned here only for a scaling comparison of parameter counts.
}

\label{tab:parameter_counts}
\end{table}

The hidden size $h=3$ was intentionally chosen to provide a comparable recurrent baseline whose parameter count increases with recurrent depth on a scale comparable to the quantum models.

\subsection{ARIMA}
\label{app:arima}

The ARIMA baseline was applied directly to the chronologically ordered precipitation series rather than to fixed \(14\) month input windows. Model order was selected using the training and validation periods while preserving their temporal ordering. We considered the grid
\begin{equation}
p\in\{0,\ldots,12\},
\qquad
d\in\{0,1\},
\qquad
q\in\{0,\ldots,6\}.
\end{equation}

Stationarity and invertibility constraints were enforced during fitting. Models were estimated using the state space implementation in \texttt{statsmodels}, with a maximum of \(500\) optimizer iterations. For each model in the grid search, the model was initially fit using the training portion of the raw precipitation record, comprising observations with indices \(0,\ldots,363\). Validation performance was then evaluated over the subsequent \(70\) observations using rolling one-step-ahead forecasts. After each forecast, the corresponding observed precipitation value was incorporated into the ARIMA state before forecasting the next month, while the fitted model coefficients remained fixed. Models were ranked according to their validation RMSE.

This procedure selected $\operatorname{ARIMA}(7,0,6)$. After model-order selection, the selected ARIMA specification was refit using all \(434\) observations preceding the test interval. Test predictions were then generated sequentially over indices \(434,\ldots,538\) using the same rolling one-step-ahead procedure. Following each test forecast, the corresponding observed precipitation value was incorporated into the model state without re-estimating the ARIMA coefficients.

Consequently, the ARIMA baseline differs from the fixed-window models in that each prediction can use the complete observed history available up to the forecast time, rather than only the preceding \(14\) months. No seasonal ARIMA terms were included.

\begin{table}[h]
\centering

\begin{tabular}{lll}
\toprule
\textbf{Model} & \textbf{Parameter} & \textbf{Value} \\
\midrule

Random forest
& Trees & \(500\) \\
& Maximum depth & \(5\) \\
& Minimum samples per leaf & \(2\) \\
& Maximum feature fraction & \(1.0\) \\
& Split criterion & Squared error \\
& Bootstrap sampling & Yes \\
& Independent runs & \(5\) \\[0.5em]

ARIMA
& Selected order & \((7,0,6)\) \\
& Candidate orders & \(182\) \\
& Seasonal terms & None \\
& Forecasting protocol & Rolling one-step \\
& Final fit observations & \(434\) \\[0.5em]

LSTM
& Input sequence & \(14\times1\) \\
& Hidden dimension & \(3\) \\
& Recurrent depth & \(D=1,\ldots,6\) \\
& Optimizer & Adam \\
& Learning rate & \(0.005\) \\
& Batch size & \(16\) \\
& Epochs & \(100\) \\
& Gradient clipping & \(1.0\) \\
& Independent runs & \(5\) \\
\bottomrule
\end{tabular}

\caption{
Summary of the classical baseline configurations.
}

\label{tab:classical-baseline-config}
\end{table}

\onecolumngrid

\end{document}